\documentclass[twocolumn]{aastex702}
\usepackage{graphics}
\usepackage{enumitem}
\usepackage{amsmath}
\usepackage{xcolor}
\usepackage{soul}
\usepackage{url}
\usepackage{color,soul}

\begin{document}

\shortauthors{Temim et al.}
\shorttitle{Dust in Cas A}

\title{JWST Resolves Dust Emission from Individual Ejecta Clumps in Cassiopeia A:\\ Dominant Grain Compositions Point to a Shell Merger in the Progenitor}

\author[0000-0001-7380-3144]{Tea Temim} 
\affiliation{Columbia Astrophysics Laboratory, Columbia University, 538 West 120th Street, New York, NY 10027, USA}
\email[show]{tea.temim@columbia.edu}

\author[0000-0001-8403-8548]{Richard G. Arendt}
\affiliation{Center for Space Sciences and Technology, University of Maryland, Baltimore County, Baltimore, MD 21250, USA}
\affiliation{Center for Research and Exploration in Space Science and Technology, NASA/GSFC, Greenbelt, MD 20771, USA}
\affiliation{Code 665, Observational Cosmology Lab, NASA Goddard Space Flight Center, Greenbelt, MD 20771, USA}
\email{Richard.G.Arendt@nasa.gov}

\author[0000-0002-0763-3885]{Dan Milisavljevic}
\affiliation{Department of Physics and Astronomy, Purdue University, 525 Northwestern Ave, West Lafayette, IN 47907, USA}
\affiliation{Institute for Physical Artificial Intelligence, Purdue University, West Lafayette, IN 47907, USA}
\email{dmilisav@purdue.edu}

\author[0000-0003-0913-4120]{Danielle Dickinson}
\affiliation{Department of Physics and Astronomy, Purdue University, 525 Northwestern Ave, West Lafayette, IN 47907, USA}
\email{dickinsd@purdue.edu}

\author[0000-0002-3362-7040]{J. Martin Laming}
\affiliation{Space Science Division, Code 7684, Naval Research Laboratory, Washington, DC 20375, USA}
\email{j.laming@nrl.navy.mil}

\author[0000-0001-9419-6355]{Ilse De Looze}
\affiliation{Department of Physics and Astronomy, Ghent University, Proeftuinstraat 86, N3, B-9000 Ghent, Belgium}
\email{ilse.delooze@ugent.be}

\author[0000-0003-2624-0056]{Christopher Fryer}
\affiliation{Center for Theoretical Astrophysics, Los Alamos National Laboratory, Los Alamos, NM 87545, USA}
\affiliation{Department of Physics, The George Washington University, Washington, DC 20052, USA}
\email{fryer@lanl.gov}

\author[0000-0002-5529-5593]{Mikako Matsuura}
\affiliation{Cardiff Hub for Astrophysical Research and Technology (CHART), School of Physics \& Astronomy, Cardiff University, The Parade, Cardiff CF24 3AA, UK}
\email{MatsuuraM@cardiff.ac.uk}

\author[0000-0003-2836-540X]{Salvatore Orlando}
\affiliation{INAF--Osservatorio Astronomico di Palermo, Piazza del Parlamento 1, 90134 Palermo, Italy}
\email{salvatore.orlando@inaf.it}

\author[0000-0002-6986-6756]{Patrick Slane}
\affiliation{Smithsonian Astrophysical Observatory, Center for Astrophysics \textbar\ Harvard \& Smithsonian, MS-3, 60 Garden Street, Cambridge, MA 02138, USA}
\email{slane@cfa.harvard.edu}

\author[0000-0001-5510-2424]{Nathan Smith}
\affiliation{Steward Observatory, University of Arizona, 933 N. Cherry Avenue, Tucson, AZ 85721, USA}
\email{nathans@as.arizona.edu}

\author[0000-0003-3643-839X]{Jeonghee Rho}
\affiliation{SETI Institute, 189 Bernardo Avenue, Suite 200, Mountain View, CA 94043, USA}
\affiliation{Department of Physics and Astronomy, Seoul National University, Gwanak-ro 1, Gwanak-gu, South Seoul, 08826, Republic of Korea}
\email{jrho@seti.org}

\author[0000-0003-2379-6518]{William P. Blair}
\affiliation{William H. Miller III Department of Physics \& Astronomy, Johns Hopkins University, 3400 N Charles St, Baltimore, MD 21218, USA}
\email{wblair@jhu.edu}

\author[0000-0003-2238-1572]{Ori D. Fox}
\affiliation{Space Telescope Science Institute, 3700 San Martin Drive, Baltimore, MD 21218, USA}
\email{ofox@stsci.edu}

\author[0000-0001-8532-3594]{Claes Fransson}
\affiliation{Department of Astronomy, Stockholm University, The Oskar Klein Centre, AlbaNova, SE-106 91 Stockholm, Sweden}
\email{claes@astro.su.se}

\author[0000-0002-0831-3330]{Hans-Thomas Janka}
\affiliation{Max Planck Institute for Astrophysics, Karl-Schwarzschild-Str.~1, 85748 Garching, Germany}
\email{thj@mpa-garching.mpg.de}

\author[0000-0002-3036-0184]{Florian Kirchschlager}
\affiliation{Department of Physics and Astronomy, Ghent University, Proeftuinstraat 86, N3, B-9000 Ghent, Belgium}
\email{florian.kirchschlager@ugent.be}

\author[0000-0003-1415-5823]{Paul P. Plucinsky}
\affiliation{Smithsonian Astrophysical Observatory, Center for Astrophysics \textbar\ Harvard \& Smithsonian, MS-3, 60 Garden Street, Cambridge, MA 02138, USA}
\email{plucinsky@cfa.harvard.edu}

\author[0000-0002-5858-6265]{Felix D. Priestley}
\affiliation{Cardiff Hub for Astrophysical Research and Technology (CHART), School of Physics \& Astronomy, Cardiff University, The Parade, Cardiff CF24 3AA, UK}
\email{PriestleyF@cardiff.ac.uk}

\author[0000-0002-7868-1622]{John C. Raymond}
\affiliation{Smithsonian Astrophysical Observatory, Center for Astrophysics \textbar\ Harvard \& Smithsonian, MS-3, 60 Garden Street, Cambridge, MA 02138, USA}
\email{jraymond@cfa.harvard.edu}

\author[0000-0002-4410-5387]{Armin Rest}
\affiliation{Space Telescope Science Institute, 3700 San Martin Drive, Baltimore, MD 21218, USA}
\affiliation{William H. Miller III Department of Physics \& Astronomy, Johns Hopkins University, 3400 N Charles St, Baltimore, MD 21218, USA}
\email{arest@stsci.edu}

\author[0000-0002-4708-4219]{Jacco Vink}
\affiliation{Anton Pannekoek Institute for Astronomy \& GRAPPA, University of Amsterdam, Science Park 904, 1098 XH Amsterdam, The Netherlands}
\affiliation{SRON Netherlands Institute for Space Research, Niels Bohrweg 4, 2333 CA Leiden, The Netherlands}
\email{j.vink@uva.nl}

\author[0009-0006-5127-8290]{Noah Zimmer}
\affiliation{Department of Physics and Astronomy, Purdue University, 525 Northwestern Ave, West Lafayette, IN 47907, USA}
\email{zimmer74@purdue.edu}



\begin{abstract}

We present a study of dust emission in the supernova remnant (SNR) Cassiopeia A (Cas A) observed with the MIRI Medium-resolution Spectrometer (MRS) on board JWST. The unprecedented spatial resolution at mid-IR wavelengths allows us to isolate dust emission from individual arcsecond-scale clumps of supernova (SN) ejecta and circumstellar material (CSM) and identify the dominant dust species in each chemically distinct region. We find that dust emission in the argon-bright reverse-shocked ejecta is dominated by SiO$_2$ grains, while the neon-bright and CO-rich ejecta region is dominated by amorphous carbon. 
Using MRS spectra of three representative regions to map similar emission properties across the SNR, we find that the SiO$_2$-associated 21~\micron\ dust emission feature is present throughout much of the reverse-shocked ejecta, while carbon dust emission is concentrated in two neon-rich crescent-shaped regions.
The spectrum of the Green Monster structure is reproduced by warm Mg$_2$SiO$_4$ dust, whose spatial distribution supports a CSM origin. 
For the ejecta, the dust compositions provide independent constraints on the abundances of refractory elements that are not fully captured by gas-phase measurements. Combining these constraints with previous ejecta line measurements, we compare the resulting abundance conditions with one-dimensional (1-D) models of SN explosion nucleosynthesis and find that they favor stellar progenitors that underwent a carbon-oxygen shell merger, adding to the growing evidence for this scenario in Cas~A.
Our study demonstrates that spatially resolved dust compositions can provide constraints on the chemical structure and properties of core-collapse explosions, with broader implications for interpreting unresolved mid-IR emission from extragalactic SNe.

\end{abstract}




\section{Introduction}

Core-collapse supernovae (SNe), the explosive deaths of massive stars, are responsible for the production and distribution of heavy elements and dust that enrich the interstellar medium (ISM) and play a crucial role in shaping the energy balance, structure, and chemical composition of galaxies. Yet many important questions remain about the nature of these explosions and their progenitors \citep[e.g.,][]{janka12, burrows21,jerkstrand26}. We still do not understand the nature and role of the progenitor system’s mass loss in determining a massive star's pre-explosion structure \citep{smith14}, how these final stages of the progenitor's evolution shape the subsequent SN explosion and its nucleosynthesis \citep[e.g.,][]{woosley02}, or how different progenitor systems produce the observed diversity of core-collapse SNe \citep[e.g.,][]{smartt09,vandyk17}. 

Young Galactic supernova remnants (SNRs) provide spatially resolved views of the metal-rich material ejected in the explosion and the circumstellar environment shaped by the pre-SN mass loss \citep[e.g.,][]{patnaude17}, making them powerful probes of the link between the properties of progenitor stars and the SNe they produce. With its unparalleled spatial resolution and sensitivity, JWST can map the ejecta and circumstellar material (CSM) structure of SNRs in unprecedented detail, as demonstrated by recent observations of the youngest Galactic core-collapse SNR Cassiopeia A \citep[Cas A, ][]{milisavljevic24}, which has an age of $\sim$350 yr \citep{fesen06} and a well-established SN type \citep[Type IIb,][]{rest08,krause08, rest11}. JWST imaging of Cas~A uncovered an intricate web-like network of oxygen-emitting ejecta gas interior to the reverse shock \citep{milisavljevic24}, likely arising from turbulent mixing driven by neutrino-heated bubbles shortly after core collapse \citep{orlando25b}. It also revealed small-scale circular holes within a structure termed the ``Green Monster'' that provided new insight about the interaction between the SN ejecta and the circumstellar shell in Cas~A \citep{vink24,delooze24,orlando25}. JWST spectroscopy confirmed the highly inhomogeneous nature of the ejecta interior \citep{milisavljevic24} and detected abundant molecular CO emission in reverse-shocked ejecta regions \citep{rho24}. In this work, we focus on the composition of dust in Cas A and what it may reveal about the chemical structure of the progenitor.

Characterizing the properties of dust grains produced in SNe provides a direct test for theoretical dust condensation models and probes the chemical conditions in which different grain species form \citep[e.g.,][]{todini01,bianchi07,nozawa10,cherchneff10,sarangi15}. Since the formation of particular dust species reflects the elemental composition of the SN ejecta, identifying the dominant grain compositions has broader implications for constraining the underlying abundance structure, particularly for refractory elements that can condense into dust.

This is especially relevant for Cas A, where growing evidence from X-ray observations suggests that the progenitor may have undergone a shell merger prior to explosion \citep{sato2025, xrism26, boccioli26b}, a phenomenon in which the oxygen-burning shell merges outward with the overlying carbon- and neon-burning shells. Shell mergers have long been predicted by one-dimensional (1-D) stellar evolution models \citep[e.g.,][]{rauscher2002,tur2007} and remain a leading uncertainty in models of late-stage massive-star evolution. Three-dimensional (3-D) simulations of shell mergers indicate that they are driven by the entrainment of material from stable into convective regions.  This erodes the stable layers that separate convective shells, making it possible for them to merge, and producing large-scale compositional mixing and asymmetries \citep{arnett_meakin2011,yadav20,andrassy2020,rizzuti2024}. The mergers can produce distinctive nucleosynthetic signatures, including enhanced abundances of odd-$Z$ elements such as P, Cl, K, and Sc, as well as an excess of oxygen-burning products such as Si and Ar relative to Ne \citep{ritter2018,roberti25}. In Cas A, several of these signatures have already been identified, including enhanced odd-Z abundances \citep{xrism26} and a high Ar/Ne ratio \citep{boccioli26b}, which is insensitive to dust condensation. An important additional line of evidence for shell-merger mixing in Cas~A comes from gas-phase Si/Mg and Ne/Mg ratios derived from X-ray observations \citep{sato2025}. Because Si and Mg can condense into dust, uncertainties in the dust composition can complicate the interpretation of these particular gas-phase ratios \citep{boccioli26b}.


The uncertainty about the dust composition in Cas~A in large part stems from the limited spatial resolution of previous mid-IR observations.
\textit{Spitzer} Infrared Spectrograph (IRS) observations of Cas~A showed that particular spectral features spatially correlate with emission lines, such as argon and neon, demonstrating that different dust grain families were forming in chemically distinct regions of the ejecta and CSM \citep{rho08,arendt14}. However, the spatial resolution of IRS ($\sim$2.5--6\arcsec) was insufficient to resolve individual ejecta knots, and each extracted spectrum required several dust components of different compositions and temperatures, making it difficult to identify the dominant compositions and cross-correlate them with precise gas chemistry. Here, we utilize the significantly higher spatial resolution of JWST to isolate individual ejecta clumps in Cas~A and identify the dust species associated with each chemically distinct region. We then use these spatially resolved dust compositions, together with gas-phase abundance information, to place new constraints on the ejecta abundance structure and evaluate the shell-merger scenario for the Cas~A progenitor.

The paper is organized as follows. Section~\ref{redux} describes the JWST Medium-resolution Spectrometer (MRS) observations and data reduction. Section~\ref{dustcomposition} presents the dust composition analysis for each MRS position. Section~\ref{dustspatial} describes the spatial distribution of dust properties across the SNR. Section~\ref{disc} summarizes the main results and discusses implications for the progenitor's pre-explosion structure. The conclusions are presented in Section~\ref{conclusion}.

\begin{figure}
\centering
\includegraphics[width=0.48\textwidth]{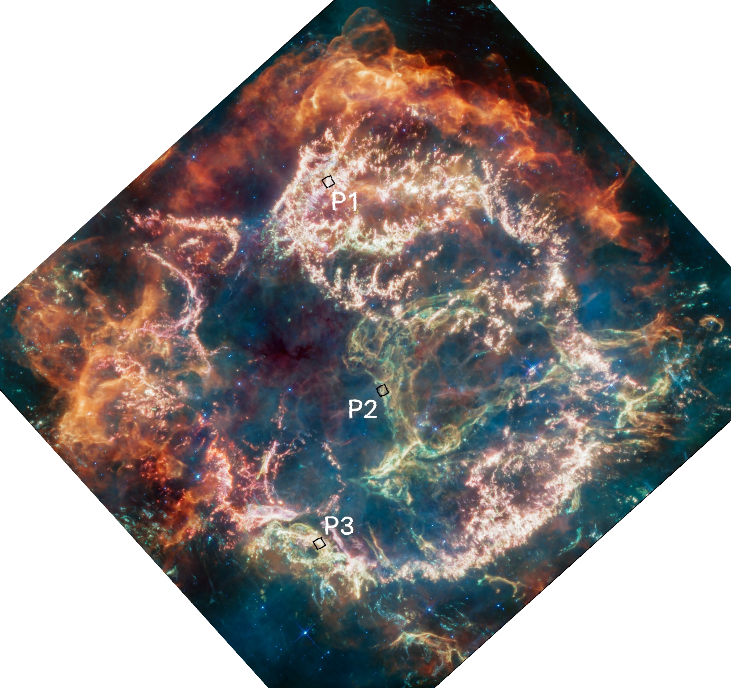} \\
\includegraphics[width=0.48\textwidth]{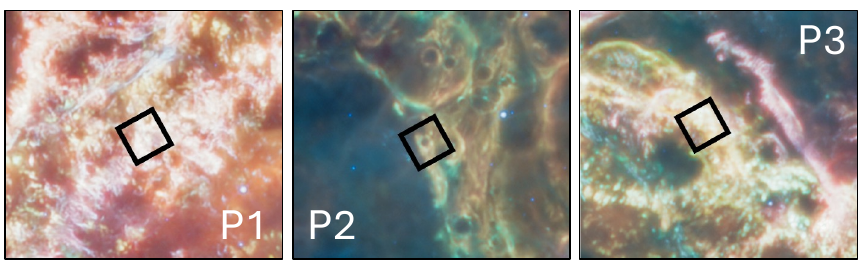}
\caption{\label{press_image} The three MRS fields of view ($\sim$~8\arcsec on a side) at positions P1, P2, and P3 are overlaid in black on the JWST press release image of Cas~A \citep{milisavljevic24}, with a zoomed-in view in the bottom panels. The three regions were intentionally selected to sample regions of distinctly different dust spectral properties previously identified from \textit{Spitzer} observations.
Image credit: NASA, ESA, CSA, Danny Milisavljevic (Purdue University), Tea Temim (Princeton University), Ilse De Looze (UGhent); Image Processing: Joseph DePasquale (STScI)}
\end{figure}

\begin{figure*}
\center
\includegraphics[width=1.0\textwidth]{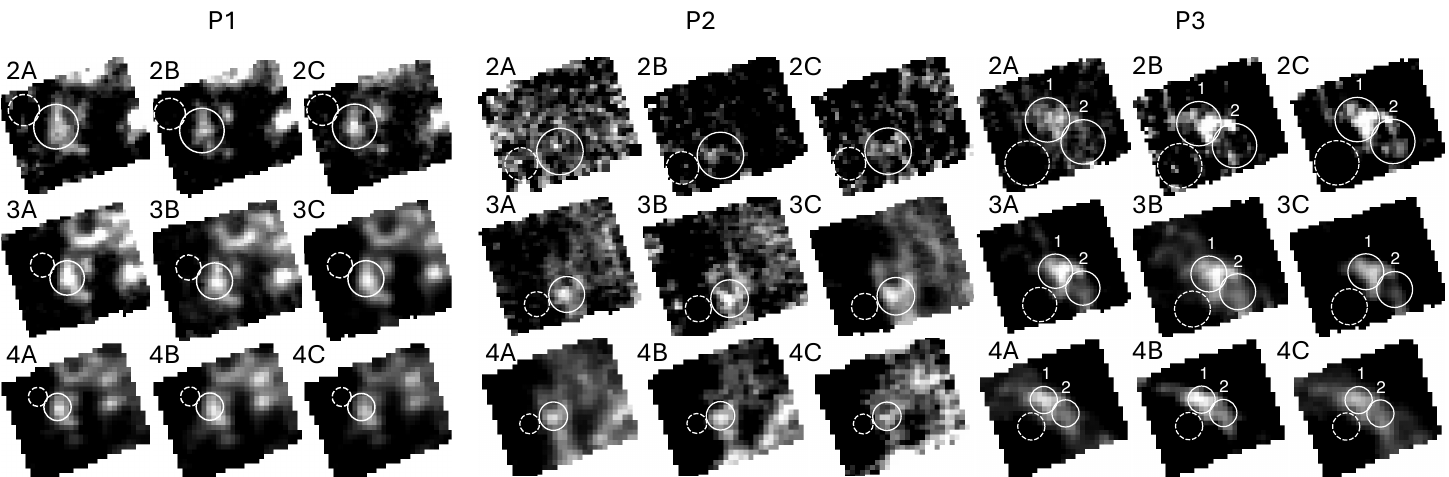}
\caption{\label{mrs_fov} Footprints of the MRS cubes at the P1, P2, and P3 positions (see Figure~\ref{press_image}) for MRS channels 2--4. The wavelength ranges covered by the channels are 2A: 7.5--8.9~$\mu$m, 2B: 8.6--10.3~$\mu$m, 2C: 10.0--11.9~$\mu$m, 3A: 11.5--13.7~$\mu$m, 3B: 13.2--15.8~$\mu$m, 3C: 15.3--18.2~$\mu$m, 4A: 17.5--21.1~$\mu$m, 4B: 20.4--24.7~$\mu$m, and 4C: 23.9--27.9~$\mu$m. The emission is shown for a narrow wavelength range within each channel that is dominated by dust emission. The white solid circles represent the spectral extraction apertures for the chosen emission knots in each region, while the white dashed lines represent the extraction apertures for the local background. The solid-circle apertures have radii of 1.0\arcsec, while the background apertures have radii of 0.7\arcsec\ at P1 and P2 and 1.0\arcsec\ at P3.. The resulting spectra are shown in Figure~\ref{spectra}.}
\end{figure*}

\begin{table*}
\centering
\begin{tabular}{lccll}
\hline
\hline
Position &
Source Coordinates (J2000) &
Background Coordinates (J2000) & F356W (MJy/sr) & F560W (MJy/sr) \\
\hline

P1 &
23:23:29.603, +58:50:31.502 &
23:23:29.788, +58:50:32.245 & 1.073 $\pm$ 0.054 & 1.932 $\pm$ 0.097 \\

P2 &
23:23:25.953, +58:48:42.512 &
23:23:26.167, +58:48:41.973 & 0.054 $\pm$ 0.003 & 0.304 $\pm$ 0.015\\

P3 (1) &
23:23:30.334, +58:47:25.407 &
23:23:30.452, +58:47:23.459 & 1.412 $\pm$ 0.071 & 12.13 $\pm$ 0.61 \\

P3 (2) &
23:23:30.128, +58:47:24.419 & 
23:23:30.452, +58:47:23.459 & 0.635 $\pm$ 0.032 & 5.969 $\pm$ 0.298 \\

\hline
\end{tabular}
\caption{\label{mrs_regions} Center coordinates for the circular source and background extraction regions used for the MIRI/MRS spectra. All source apertures have radii of 1\arcsec, while the background apertures are 0.7\arcsec\ in radius for P1 and P2, and 1\arcsec for P3. The source surface brightness values in MJy/sr using the same source and background apertures are listed for the NIRCam F356W and MIRI F560W filters.}
\end{table*}

\section{Observations and Data Reduction} \label{redux}

The MIRI Medium-Resolution Spectrometer \citep[MRS;][]{argyriou23} spectroscopic observations were obtained as part of the JWST Cycle 1 GO program 1947 (PI: Milisavljevic) in November 2022 and first presented in \citet{milisavljevic24}. Here we analyze three MRS positions centered on J2000 coordinates 23:23:29.5212, +58:50:31.340 (Position 1, hereafter P1), 23:23:25.9879, +58:48:43.460 (Position 2, hereafter P2), and 23:23:30.3311, +58:47:25.020 (Position 3, hereafter P3). The locations of these three positions overlaid on the JWST image of Cas~A are shown in Figure~\ref{press_image}. They were selected in part to sample the primary dust families previously identified in the Spitzer IRS maps of Cas~A \citep{rho08,arendt14}. P1 and P3 target emission knots in the contrasting regions of the reverse-shocked ejecta \citep{milisavljevic24}, while P2 samples one of the emission rings in the ``Green Monster" region attributed to CSM on the near side of Cas~A \citep{vink24, delooze24, orlando25}. 

The observations were carried out using all four MRS channels and all three grating settings (SHORT, MEDIUM, and LONG), therefore covering the full 4.9-27.9~$\micron$ MRS wavelength range. For each grating, the exposure parameters consisted of 50 groups and one integration per dither, using a 4-point dither pattern optimized for extended sources. This resulted in a total exposure time of 555~s.
The data reduction and creation of MRS data cubes were carried out using version $\texttt{1.18.0}$ of the JWST calibration pipeline and the $\textit{jwst\_1414.pmap}$ CRDS context file. Each channel and grating combination was processed separately, resulting in 12 individual cubes for each of the three positions. Although a dedicated MRS background was also obtained for program 1947, it was not subtracted from the MRS cubes for our analysis since we opted to use a local background extracted within the MRS cube for each position.

\begin{figure*}
\center
\includegraphics[width=0.49\textwidth]{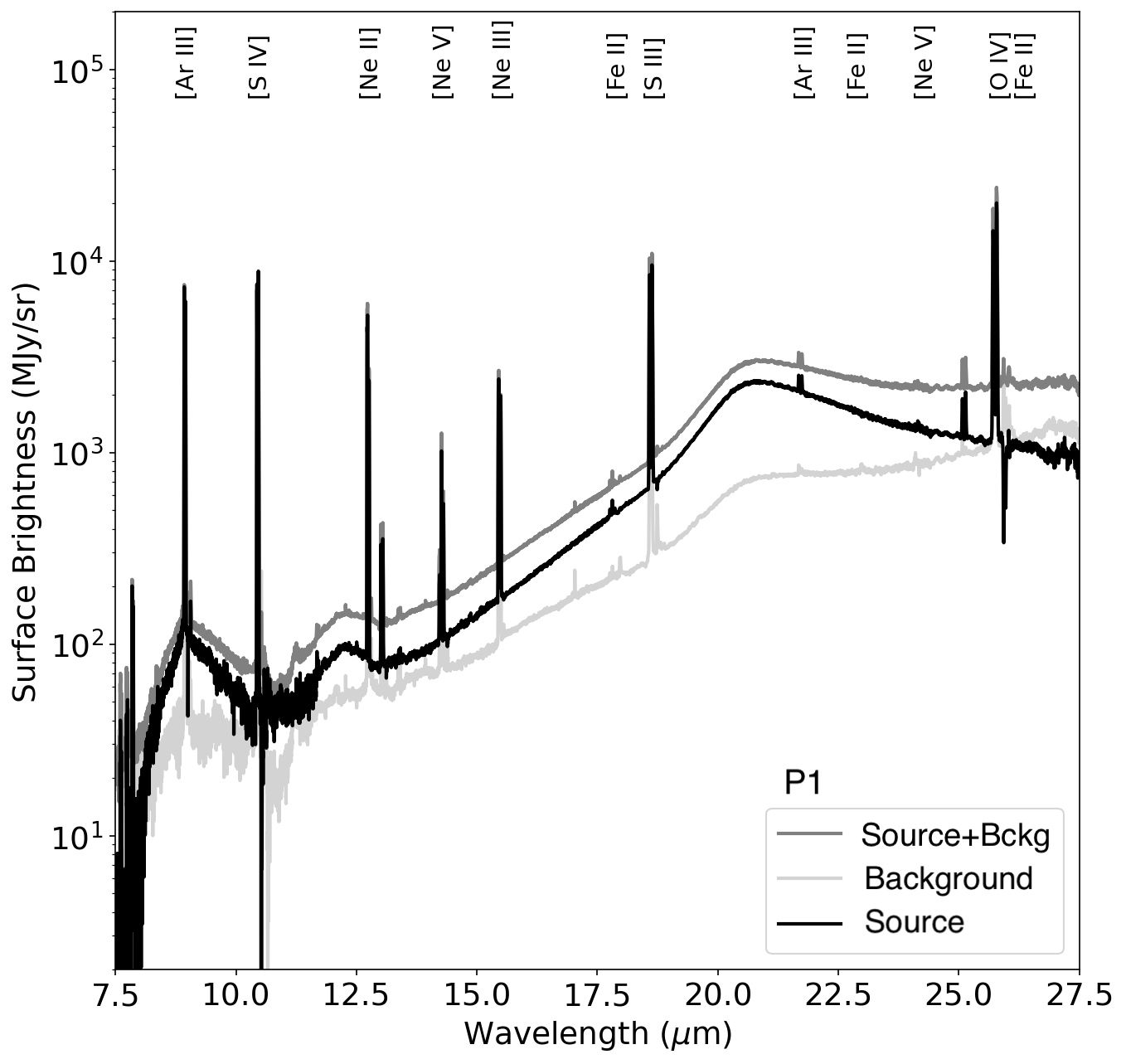}
\includegraphics[width=0.49\textwidth]{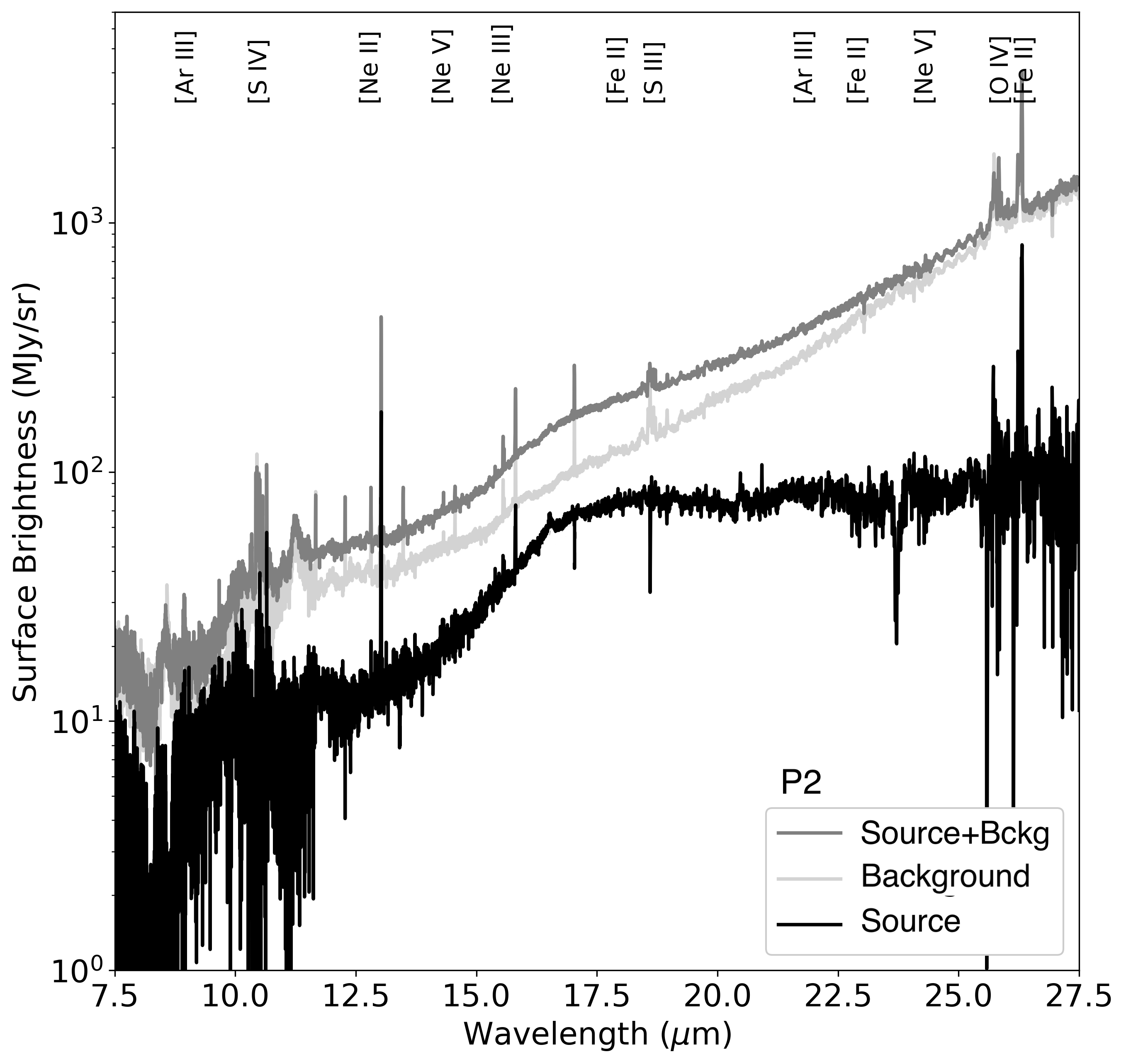} \\
\includegraphics[width=0.49\textwidth]{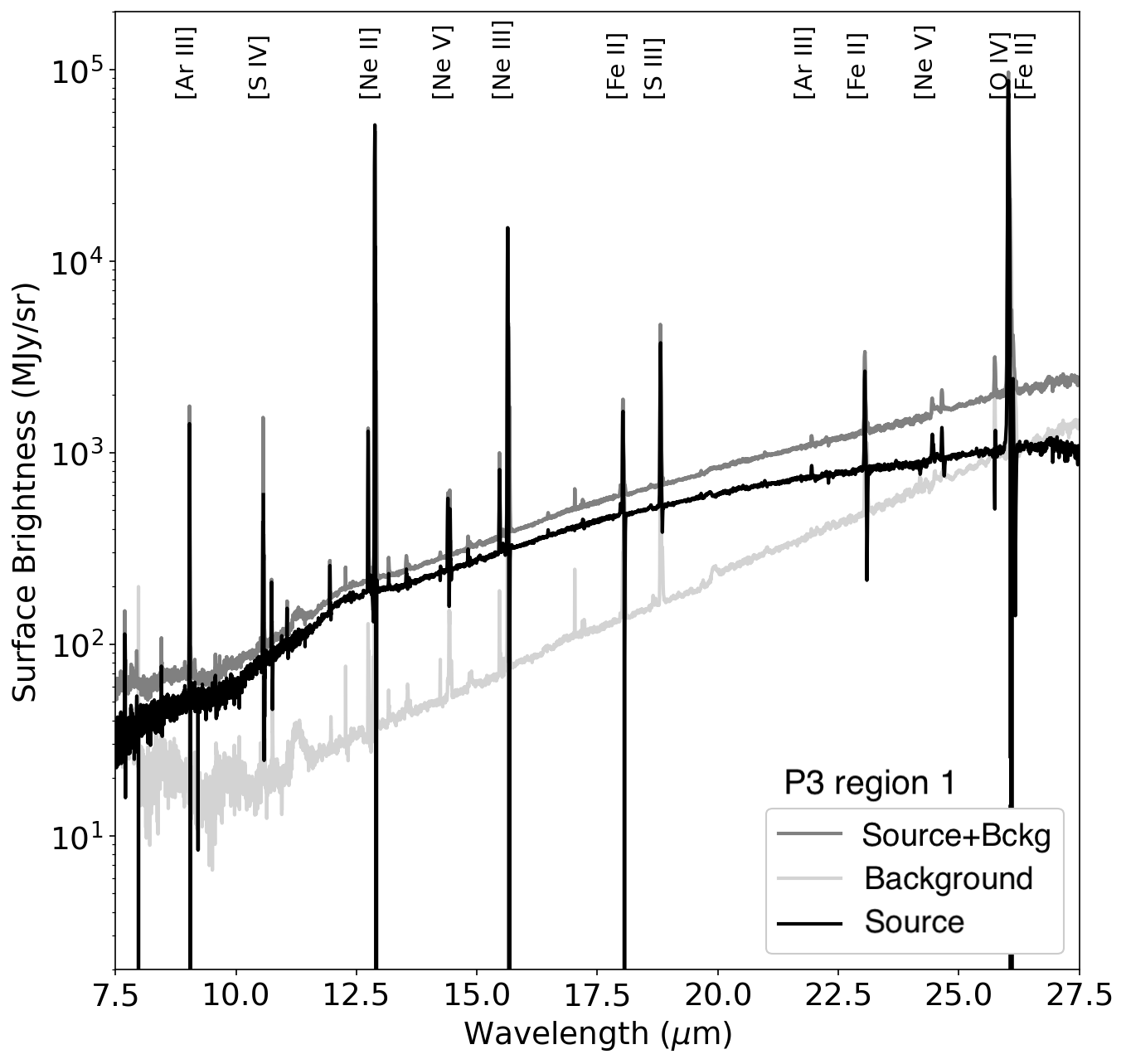}
\includegraphics[width=0.49\textwidth]{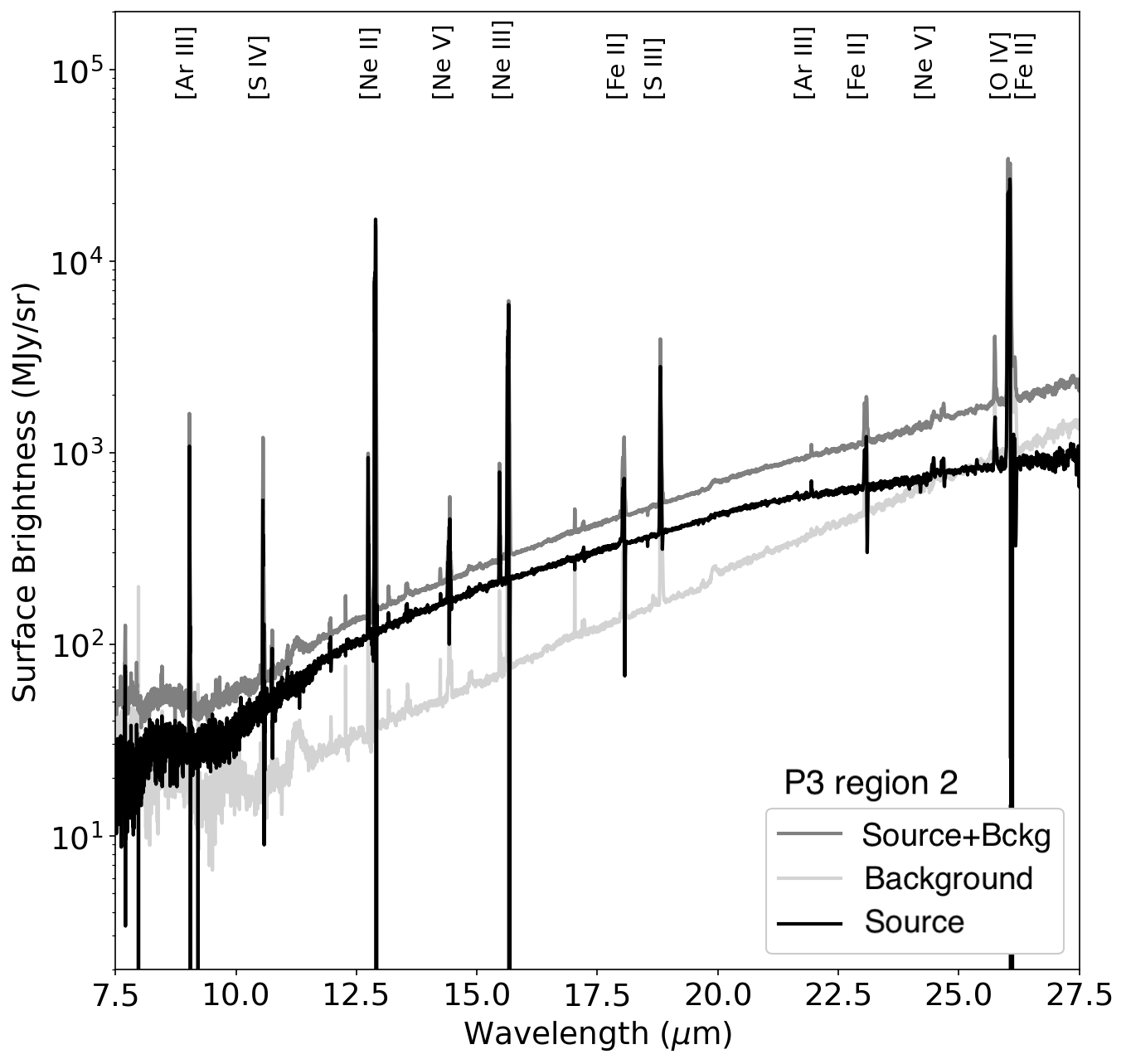}
\caption{\label{spectra} MRS spectra extracted from the regions shown in Figure~\ref{mrs_fov}. The background-unsubtracted spectra extracted from the solid white circles in Figure~\ref{mrs_fov} are shown in dark gray, the spectra extracted from the local background regions (dashed white circles in Figure~\ref{mrs_fov}) are shown in light gray, and the local-background-subtracted spectra are shown in black. The spectral ``bump" at $\sim$ 12.5~\micron\ in P3 region 1 is likely a light-scattering artifact caused by the bright [\ion{Ne}{2}] line at 12.8~\micron.}
\end{figure*}

\begin{figure*}
\center
\includegraphics[width=0.46\textwidth]{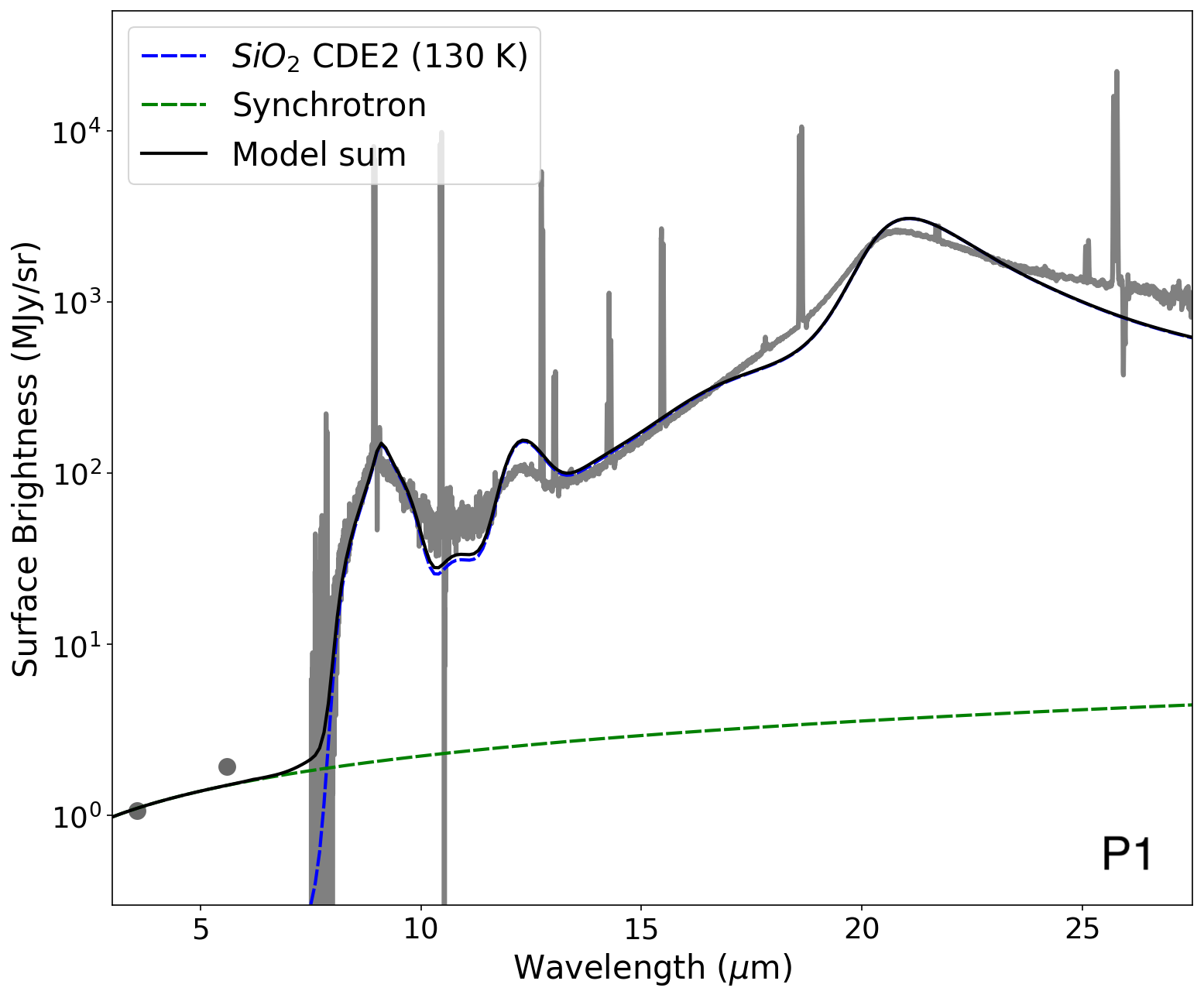}
\includegraphics[width=0.46\textwidth]{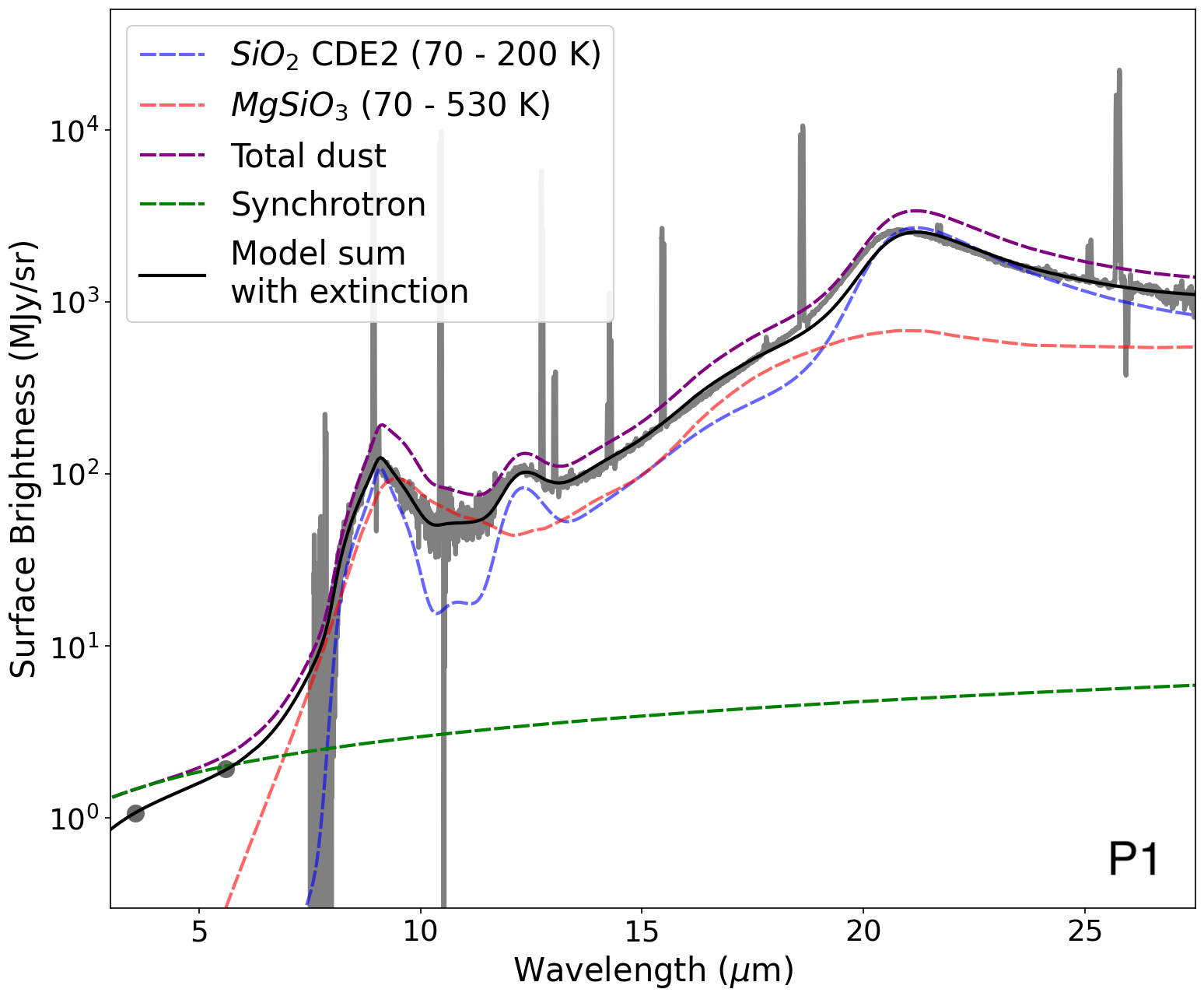}\\
\vspace{-4mm}
\includegraphics[width=0.46\textwidth]{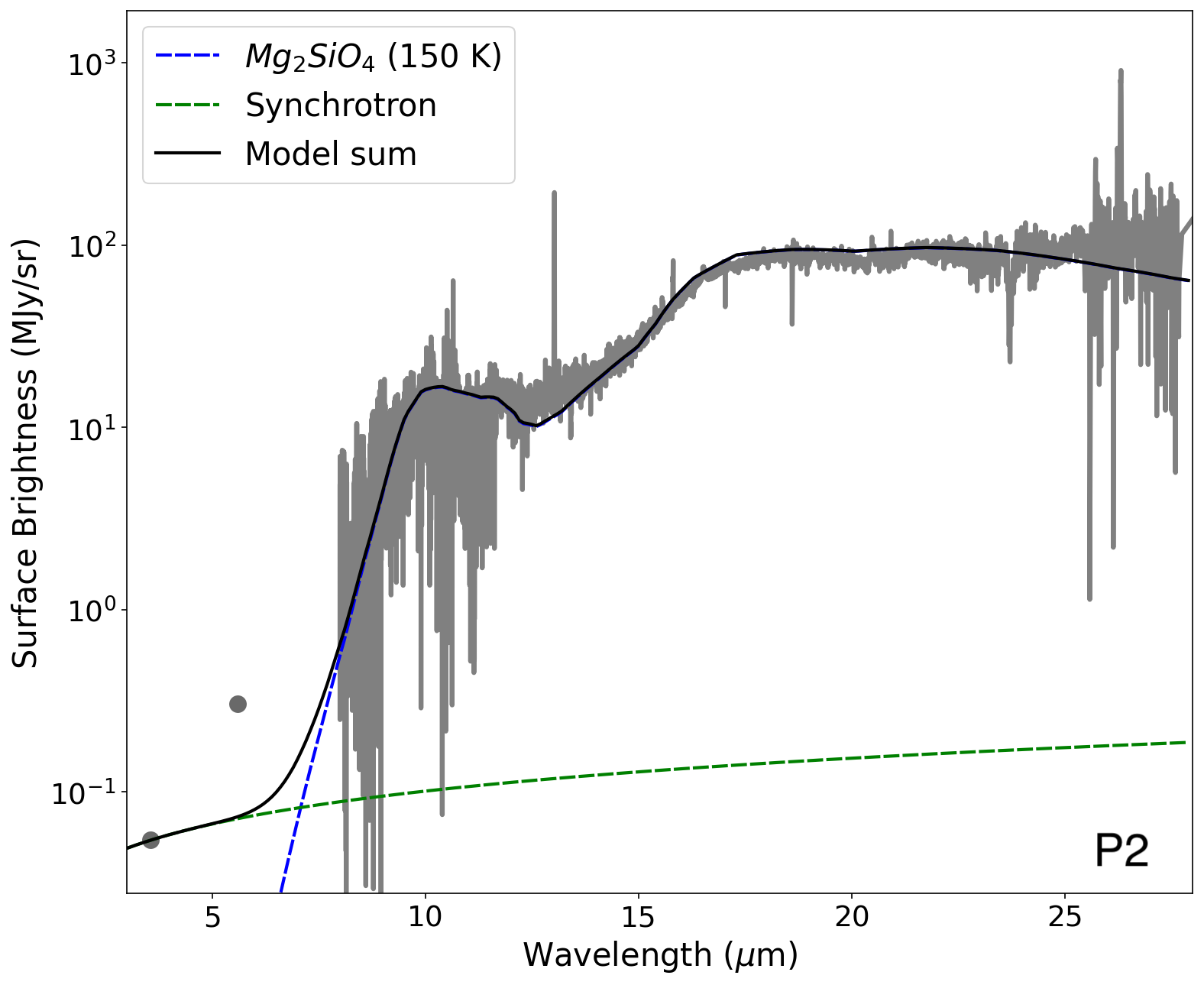}
\includegraphics[width=0.46\textwidth]{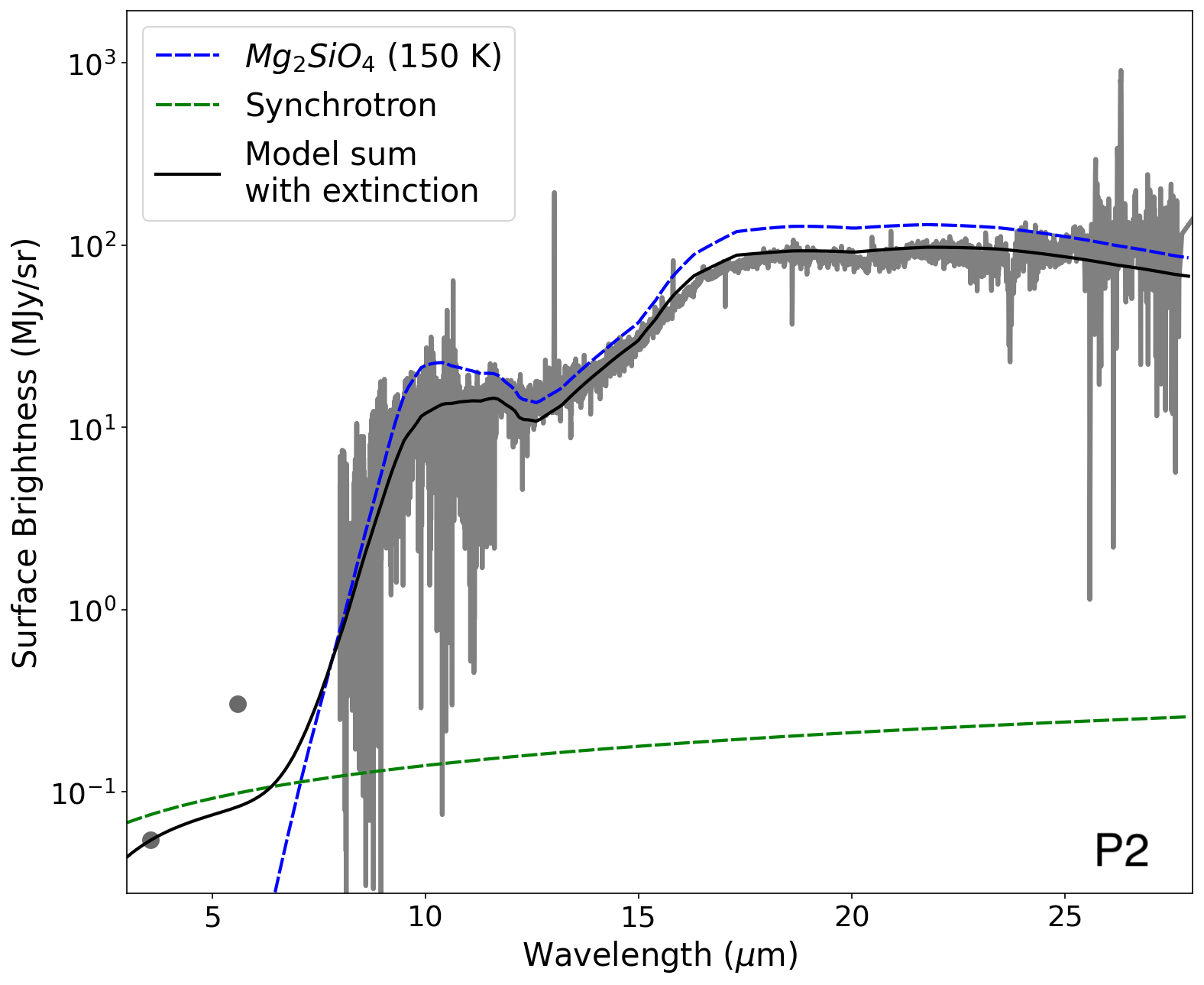}\\
\vspace{-4mm}
\includegraphics[width=0.46\textwidth]{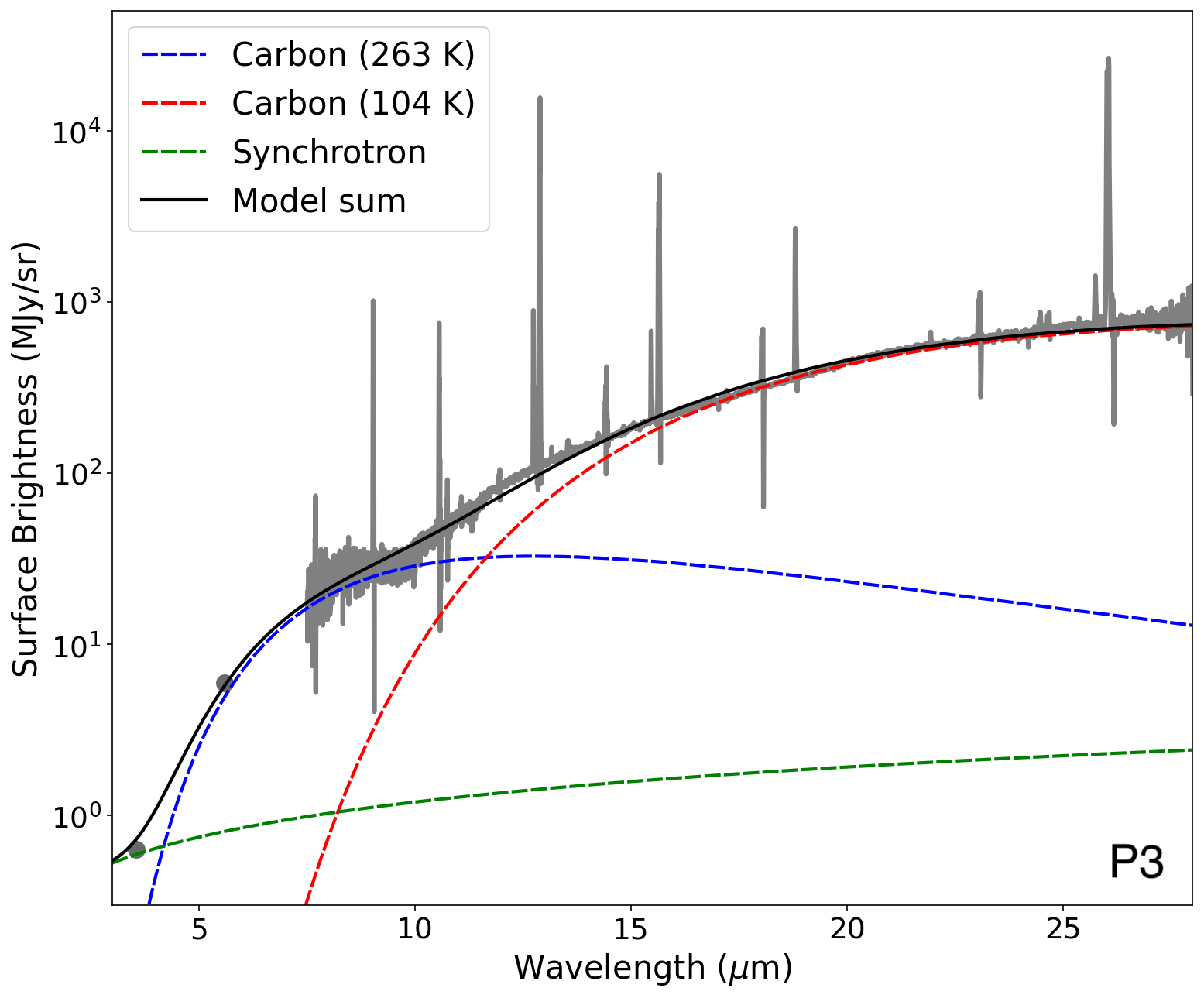}
\includegraphics[width=0.46\textwidth]{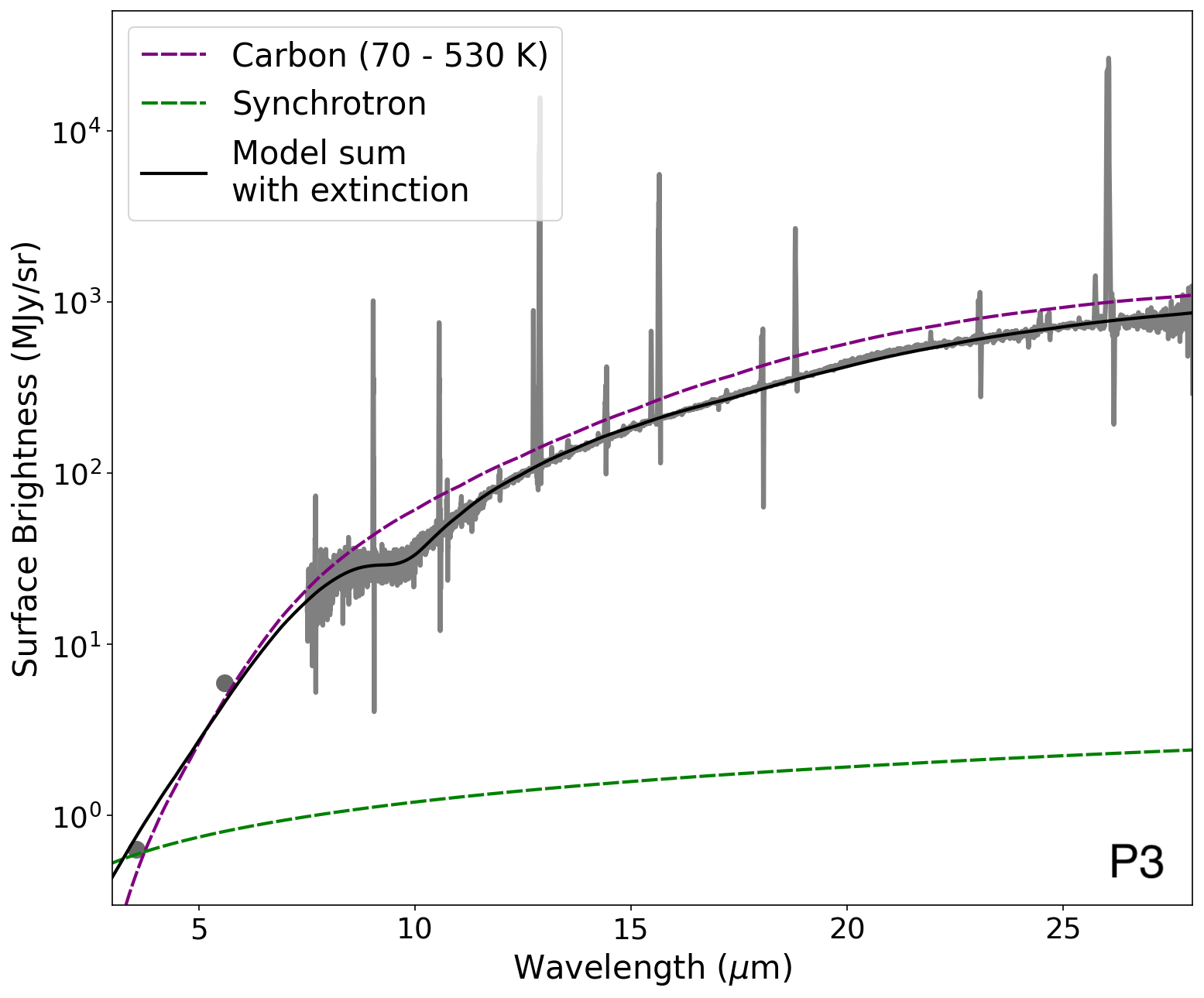}
\caption{\label{dustfits} Spectral fits corresponding to the models summarized in Table~\ref{fit_results_table}. \textbf{Top row:} Fit to the spectrum for position P1, showing a single-temperature component of SiO$_2$ on the left and a continuous-temperature mixture of SiO$_2$ and MgSiO$_3$ with foreground extinction on the right. \textbf{Middle row:} Fit to the spectrum for position P2, showing single-temperature Mg$_2$SiO$_4$ dust without (left) and with foreground extinction (right). \textbf{Bottom row:} Fit to the spectrum for position P3 (region 2), showing the two-temperature carbon dust fit without foreground extinction on the left, and a continuous-temperature distribution of carbon dust with foreground extinction on the right. Gray circles represent the F356W and F560W flux densities from the same extraction apertures, while the green dashed curve represents synchrotron emission with an assumed power-law spectral index of -0.68 for regions P1 and P3, and -0.6 for region P2 \citep{domcek21}.}
\end{figure*}

\subsection{Spectral Extraction} \label{mrs}

For each position, spectra were extracted from compact emission structures within the MRS field of view. The footprints for channels 2--4 and the chosen extraction apertures are shown in Figure~\ref{mrs_fov}. 
The goal was to determine the dust composition within individual CSM and ejecta clumps, so we aimed to minimize the contamination from unrelated emission along the line of sight, as well as diffuse SNR emission surrounding the clumps. 
For this reason, we chose to subtract background spectra extracted adjacent to each clump of emission. We excluded MRS channel 1 covering the 5.0--7.5~$\micron$ spectral range in our analysis because its smaller field of view of 3.2\arcsec$\times$3.7\arcsec\ did not contain an adequate off-source region for local background extraction and because at these shorter wavelengths the continuum becomes too faint to provide any useful constraints on modeling the dust emission.

The extraction apertures for the source and background regions are shown in Figure~\ref{mrs_fov}. All source apertures centered on the emission clumps were chosen to be 1.0\arcsec\ in radius to minimize the effects of the larger PSF size at longer wavelengths. The aperture size for the more diffuse background emission was chosen to be 1.0\arcsec\ in radius at P3, but a smaller radius of 0.7\arcsec\ was required at P1 and P2 because larger clump-free emission regions were not available within the field of view.

For each position, the source and local background spectra were extracted from each of the nine retained MRS cubes separately (three grating settings for each of channels 2-4). The background spectra were subtracted from the source spectra for each order, and the resulting spectra stitched together using multiplicative factors so that overlapping spectral regions across orders matched in amplitude. These adjustments ranged from 0 to 10\%. The resulting background-subtracted spectrum for each region is shown in black in Figure~\ref{spectra}. We also show the spectra before background subtraction (dark gray) and the background spectra (light gray) that were stitched in a similar way for comparison. The background emission in our case includes thermal emission from the telescope, zodiacal light, line-of-sight ISM, and diffuse CSM and SN ejecta emission in the immediate vicinity of each clump.
Because the local background emission contains emission lines at velocities that differ from those in the chosen clumps, our background-subtracted source spectra contain negative residual lines in addition to emission lines originating from the clump.

All background-subtracted source spectra show significant emission from dust, as well as prominent emission lines from Ar, Ne, S, and other species, particularly at positions P1 and P3 that sample the reverse-shocked ejecta. In the MRS observations of the Crab Nebula, extremely bright emission lines caused light scattering in the detector across slices and spectral channels, producing artifacts in the spectrum that appeared as broad emission features \citep{temim24}. Given the significant brightness of some emission lines in the Cas~A spectra, we inspected detector-level data for these scattering effects and found that the particularly bright [\ion{Ne}{2}] line at 12.8~$\micron$ produces a broad bump around 12.5~\micron\ in region 1 of the P3 spectrum. We extracted a source spectrum from an adjacent region in P3 (region 2 in Figure~\ref{mrs_fov}) away from the brightest Ne emission and found that the 12.5~\micron\ bump disappears (see the last panel of Figure~\ref{spectra}). We therefore used the P3 region 2 spectrum for the rest of our analysis. We did not find evidence of these scattering effects at positions P1 and P2.

\section{Dust Composition} \label{dustcomposition} 

The MRS spectra extracted at positions P1, P2, and P3 are shown in Figure~\ref{spectra}, and all exhibit significant emission from warm dust, but with spectral shapes that differ considerably between the three regions. In this section, we characterize the dominant dust composition at each position by fitting the mid-IR emission with dust models of varying grain compositions and temperature distributions.

For each position, we fitted the line-emission-subtracted 7.5--28~\micron\ spectrum and also included the NIRCam F356W and MIRI F560W \citep{milisavljevic24} photometry computed from the same extraction regions (see Table~\ref{mrs_regions}). We first attempted to fit each spectrum with a single dust composition plus a synchrotron component with a fixed synchrotron spectral index of -0.68 for regions P1 and P3 and -0.60 for region P2 \citep{domcek21,rho24}. 
Where possible, we also included a more realistic continuous distribution of grain temperatures, as would be expected for a population of grains with a range of sizes, and applied line-of-sight extinction using the G23 model \citep{gordon09,fitzpatrick19,gordon21,decleir22} within the dust extinction Python package of \citet{gordon24}.
The visual extinction A$_V$ from ISM dust projected across Cas~A is shown to range from A$_V$ $\sim$ 7 to A$_V$ $\gtrsim$ 15 \citep{delooze17, domcek21}, which would noticeably impact the spectral shape in the mid-IR. In our fits, we let the A$_V$ value vary within this range. Below we describe the dust composition choices and fitting results for each MRS-sampled position.

\begin{table*}
\centering
\begin{tabular*}{0.92\textwidth}{@{\extracolsep{\fill}} lllccc @{}}
\hline
\hline
Position & Model Description & Dust &
$T_{\rm dust}$ &
$M_{\rm dust}$ &
$A_V$ 
\\
 &
 &
Composition &
(K) &
($M_\odot$) &
(mag) 
\\
\hline

P1 &
Single-temperature &
SiO$_2$ &
$130.2 \pm 0.2$ &
$1.3\times10^{-7}$ & 
\\[1ex]

&
Continuous-temperature with extinction &
SiO$_2$ &
$70$--$200$ ($\alpha_T$=-8.7) &
$5.5\times10^{-7}$ &
$7.7 \pm 0.1 $
\\

&
 &
MgSiO$_3$ &
$70$--$530$ ($\alpha_T$=-7.8) &
$3.8\times10^{-7}$ &
\\[1ex]
\hline

P2 &
Single-temperature  &
Mg$_2$SiO$_4$ &
$150.7 \pm 0.3$ &
$6.0\times10^{-9}$ &
\\[1ex]


&
Single-temperature with extinction &
Mg$_2$SiO$_4$ &
$150.4 \pm 0.3$ &
$9.6\times10^{-9}$ &
$>11.9$ 
\\[1ex]
\hline

P3 &
Two-temperature &
Carbon &
$262.7 \pm 0.7$ &
$6.0\times10^{-10}$ &
\\

 &
 &
Carbon &
$104.4 \pm 0.1$ &
$7.5\times10^{-7}$ &
\\[1ex]

&
Continuous-temperature with extinction &
Carbon &
$70$--$530$ ($\alpha_T$=-9.3) &
$5.6\times10^{-6}$ &
$ 8.1 \pm 0.1 $ 
\\
\hline
\end{tabular*}
\caption{\label{fit_results_table} Best-fit dust model parameters for each region with and without foreground extinction. Statistical 1-$\sigma$ uncertainty on the dust masses is $\sim$1\%, but the total dust masses for the continuous temperature models are highly dependent on the chosen temperature distribution and are not constrained by these models. Similarly, the uncertainties on A$_V$ do not include any systematic uncertainties associated with the choice of optical constants, which may be substantially larger. The single- and two-temperature masses are good approximations of the minimum dust mass that produces the mid-IR emission. The parameter $\alpha_T$ represents the best-fit power-law index for the dust temperature distribution. Based on \citet{domcek21}, the synchrotron spectral index is assumed to be -0.68 for P1 and P3, and -0.6 for P2.}
\end{table*}

\subsection{Position P1}

The spectrum at P1, which samples reverse-shocked ejecta with particularly bright argon emission, is characterized by broad dust emission features at 9, 12, and 21 \micron. \citet{arendt14} attributed the 9 and 21 \micron\ features to Mg$_{0.7}$SiO${_{2.7}}$ grains characterized by a low magnesium-to-silicon ratio of $<$ 0.5, and noted that the 12 \micron\ feature could arise from a continuous distribution of ellipsoids (CDE) of SiO$_2$ or SiC. The CDE approximation was necessary to adequately broaden and shift the dust emission features. \citet{rho18} suggested that both the 9 and 21 \micron\ features originate from CDE SiO$_2$ grains, but required additional components of Mg-silicates, carbon, and SiC dust to reproduce the overall shape of the spectrum. 

The same three dust emission features are present in the spectrum of newly condensed SN dust in the supernova remnant G54.1+0.3 \citep{temim10,temim17,rho18}, suggesting that they may arise from a single grain composition or that the different grain species responsible for the features are commonly found in the same environment.

Given that the JWST data allow us to sample a very localized clump of ejecta, we first attempted to fit the P1 spectrum by a single-temperature component of SiO$_2$ dust \citep{henning97} with a CDE2 grain shape distribution considered to be more realistic due to a lower fraction of extreme grain shapes \citep{ossenkopf1992, draine21}. The best-fit model is shown in the top-left panel of Figure~\ref{dustfits} with the parameters summarized in Table~\ref{fit_results_table}. The SiO$_2$ CDE2 grains emitting at a single temperature of $\sim$~130~K reproduce the three broad emission features surprisingly well, though some spectral regions remain poorly matched.

We attempted to improve the fit by including the expected line-of-sight extinction and allowing for a distribution of grain temperatures that would naturally result from the heating of grains of different sizes. We found that including an additional MgSiO$_3$ \citep{jager03} component significantly improves the fit to the spectrum. While MgSiO$_3$ may not be a unique secondary composition, dust condensation models applied to Cas~A show that it is expected to spatially coincide with SiO$_2$ \citep{nozawa10}. For both grain compositions, the minimum and maximum grain temperature limits were set to 70 K and 530 K, respectively, and the temperature was assumed to have a power-law distribution with an index of $\alpha_T$ that was allowed to vary. We show the best fit in the top right panel of Figure~\ref{dustfits}. The MRS spectrum is well reproduced by this model at all wavelengths except the short-wavelength side of the 21~\micron\ feature, but we note that the best-fit dust masses in Table~\ref{fit_results_table} are highly dependent on the chosen temperature distributions and, therefore, are not uniquely constrained in the model. The total dust mass in continuous-temperature models is heavily weighted by the cold dust grain contribution whose minimum temperature is not well constrained by the MRS data that only extend to $\sim$28~$\micron$. The dust mass for the single-temperature fit in Table~\ref{fit_results_table} represents a lower limit on the dust mass in the clump.
Nevertheless, we demonstrate here that SiO$_2$ grains provide the distinctive spectral features in this ejecta clump and comprise a significant fraction of the mid-IR emission in P1, regardless of the exact composition of the secondary dust component.

\subsection{Position P2} \label{p2dust}

Position P2 is centered on one of the rings within the Green Monster structure in Cas~A (see Figure~\ref{press_image}). Based on the properties of the spatially coinciding X-ray emission \citep{vink24} and analysis of IR emission lines detected with JWST \citep{delooze24}, this structure was attributed to CSM emission on the near side of Cas~A. Its spectrum is characterized by broad bumps at 10 and 18~\micron.
This region is referred to as the ``South Spot" in \citet{arendt14}, and its \textit{Spitzer} IRS spectrum was fitted by a mixture of MgFeSiO$_4$ and amorphous carbon grains. 

We attempted to reproduce the MRS spectrum using a single temperature component for each of the following grain compositions: MgSiO$_3$, Mg$_2$SiO$_4$, Mg$_{2.4}$SiO$_{4.4}$, and MgFeSiO$_4$ from \citet{jager03}, silicates from \citet{ossenkopf1992}, and astrodust from \citet{hensley23}. 
We find that the Mg$_2$SiO$_4$ grains at 150~K provide the best fit to the MRS spectrum at P2, shown in the left-middle panel of Figure~\ref{dustfits}. In the right-middle panel of the same figure, we show that a similar fit is achieved with the same composition, but with line-of-sight extinction of A$_V$ $>$ 11.9. An A$_V$ value that is larger than for positions P1 and P3 is consistent with \citet{delooze17}, but its absolute value likely has additional systematic uncertainties associated with the uncertainties in the grain optical constants that are not accounted for here. The parameters for both fits are listed in Table~\ref{fit_results_table}. For position P2, applying a wider range of grain temperatures did not noticeably improve the fit. The fits for the other compositions we tested are summarized in the Appendix.

\subsection{Position P3}

\begin{figure*}
\center
\includegraphics[width=0.92\textwidth]{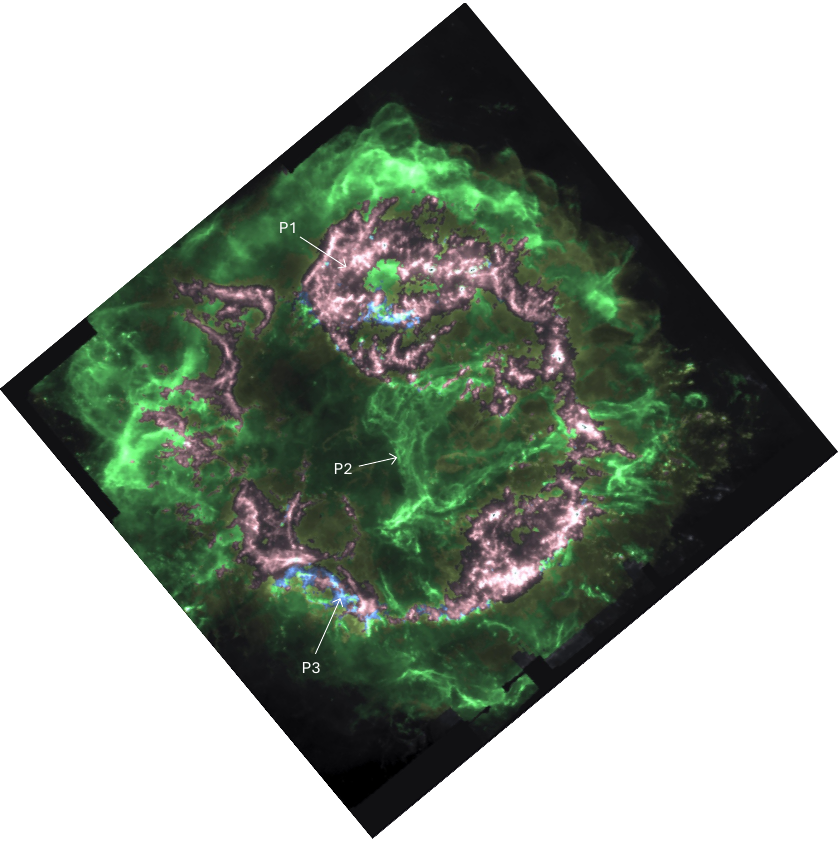}
\caption{\label{dust3color} The MIRI F2100W image of Cas~A, colored to show regions with SEDs similar to the three MRS-sampled dust families. The spatial distribution of the P2-like emission associated with Mg$_2$SiO$_4$ CSM dust is shown in green, P1-like emission associated with SiO$_2$ dust in magenta, and P3-like Ne-rich emission associated with carbon dust is shown in blue. The P1, P2, and P3 regions are indicated by the white arrows. The faint emission in brown does not show SED features characteristic of the MRS-sampled positions (no strong excess in F1800W and F2100W filters from dust features and no strong F1280W excess from Ne), and may represent cooler CSM dust or ejecta emission from lines and/or dust. The details of the spatial decomposition method for the dust components are described in Section~\ref{dustdecomposition}.}
\end{figure*}

Position P3 samples reverse-shocked ejecta with very bright neon emission relative to P1. In contrast to P1, the shape of the spectrum is smooth and featureless, with the exception of a small dip in emission around 10~\micron. \citet{arendt14} attributed this neon-correlated dust family to a mixture of Al$_2$O$_3$ and amorphous carbon grains. The MRS spectrum of this small-scale ejecta region is spectrally simpler and far less blended than the \textit{Spitzer} IRS spectrum that sampled a much larger spatial area. Its shape requires grains with a featureless emission spectrum. While both amorphous carbon and iron grains provide such a featureless spectrum, the much lower emissivity of iron leads to an implausibly high iron dust mass. To reproduce the spectrum of the individual clump, more than 0.003~M$_{\odot}$ of iron dust would be required and a minimum of $\sim$~0.3~M$_{\odot}$ total within the blue regions of Figure~\ref{dust3color}, more than the total estimated mass of Fe in Cas~A \citep{hwang12, laming20}. We therefore find that amorphous carbon is the most plausible composition at P3. 

The bottom row of Figure~\ref{dustfits} shows the best-fit models using a carbon dust composition \citep{jaeger98}. A single dust temperature could not reproduce the entire spectrum, so we first modeled the emission using two temperature components (bottom-left panel). In the bottom-right panel of Figure~\ref{dustfits}, we show that the spectrum at P3 is very well described by a continuous temperature distribution of amorphous carbon grains with the dip at 10~\micron\ naturally reproduced by silicate absorption from line-of-sight extinction. The best fit value of the visual extinction A$_V$=8.1$\pm$0.1 is consistent with the A$_V$ estimate of \citet{delooze17}. The quoted uncertainty on A$_V$ is the 1-$\sigma$ statistical uncertainty from the fit assuming the chosen optical constants. The parameters for each P3 model are listed in Table~\ref{fit_results_table}. Again, we note that the dust mass from the two-temperature fit provides a lower limit on the dust mass in the ejecta clump, whereas the mass from the continuous-temperature fit is not constrained due to its sensitivity to the chosen temperature distribution.

\begin{table*}
\centering
\setlength{\tabcolsep}{8pt}
\begin{tabular}{lllcc}
\hline
\hline
Position & Observed Property & Inferred Abundance Condition &
Standard &
Shell-Merger
\\
&
&
&
Progenitor &
Progenitor
\\
\hline

P1 &
Formation of SiO$_2$ dust & Si/Mg abundance ratio $>$ 1 &
No  &
Yes 
\\

 &
Argon-rich relative to P3 & Enhanced Ar/Ne in Si-rich ejecta &
No &
Yes
\\[1ex]

P3 &
Formation of carbon dust & Carbon-rich conditions favoring C/O $>$ 1 &
Yes  &
Yes 
\\

 &
 Ne abundance at P3 $\gtrsim$ P1 & High Ne abundance in C-rich ejecta &
No  & 
Yes 
\\

\hline
\end{tabular}
\caption{\label{obs_properties}Comparison between the observed ejecta properties in regions P1 and P3 and expectations from standard progenitor models and shell-merger scenarios. The abundance conditions inferred from the dust compositions are qualitative rather than direct abundance measurements, adopting Si/Mg $>$ 1 as an abundance condition favorable to SiO$_2$ formation, and C/O $>$ 1 as a condition favorable to carbon dust formation after the available oxygen is locked up in CO.}
\end{table*}

\begin{figure}
\center
\includegraphics[width=0.4\textwidth]{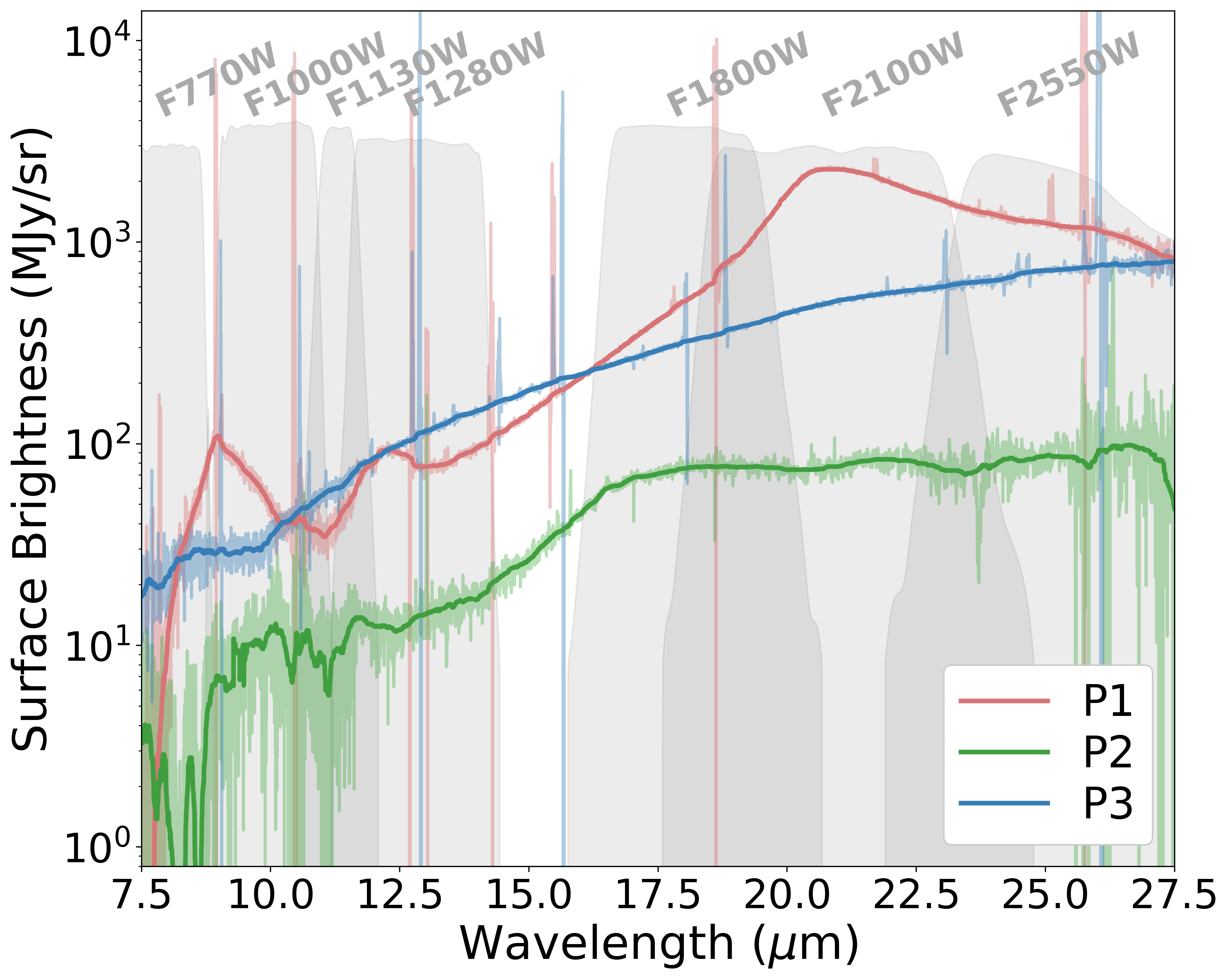} \\
\includegraphics[width=0.4\textwidth]{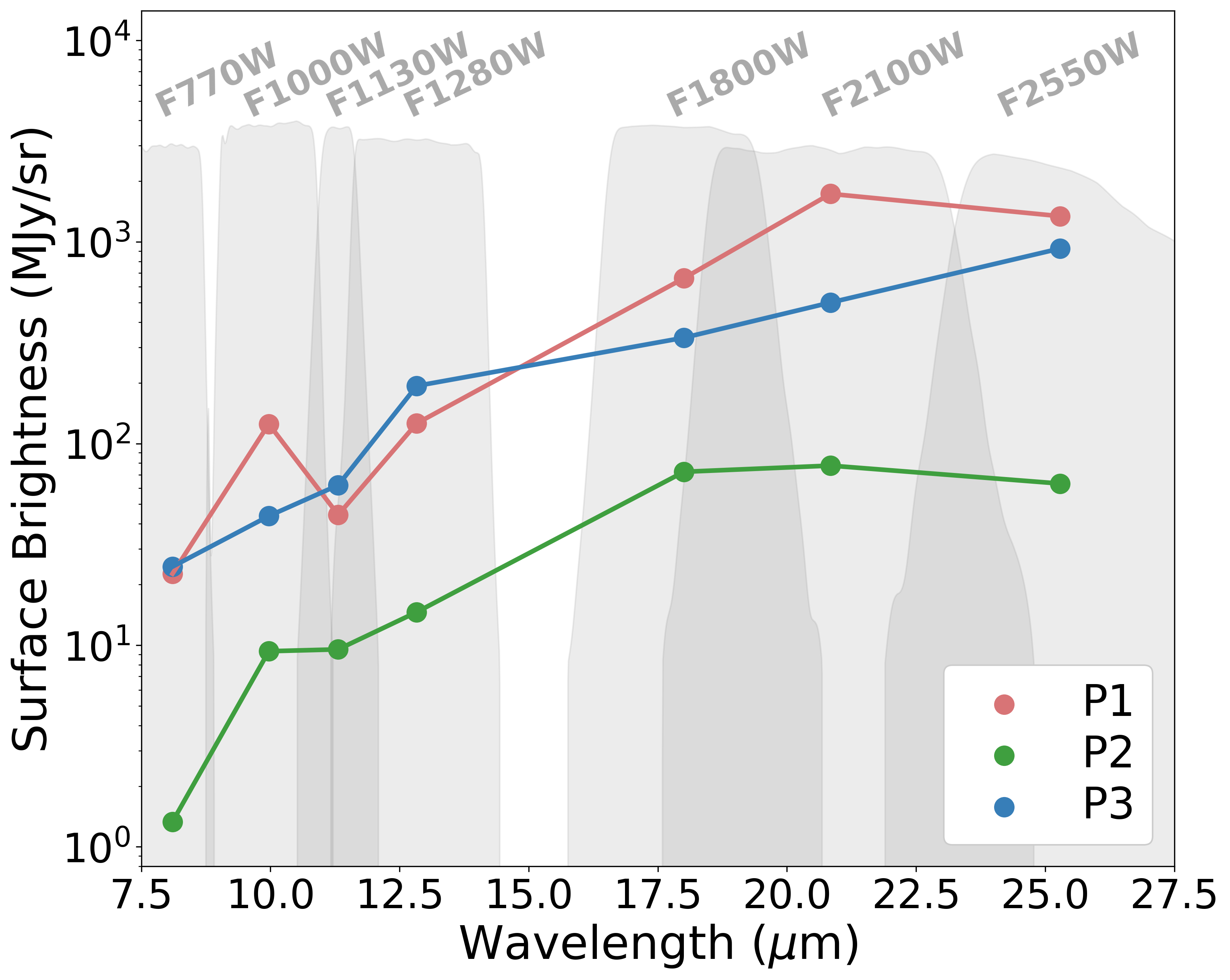}
\caption{\label{dustseds} \textbf{Top:} Dust emission spectra from the three MRS regions overlaid with the MIRI imaging filter bandpass curves in gray. \textbf{Bottom:} SEDs derived by integrating the MRS spectra in the top plot over the MIRI imaging filter bandpass curves. The SED in region P1 is characterized by an excess in the F2100W filter, P2 exhibits relatively enhanced F1800W emission, and P3 shows a smooth rising SED with an excess at F1280W from the [\ion{Ne}{2}] line.}
\end{figure}

\begin{figure}
\center
\includegraphics[width=0.45\textwidth]{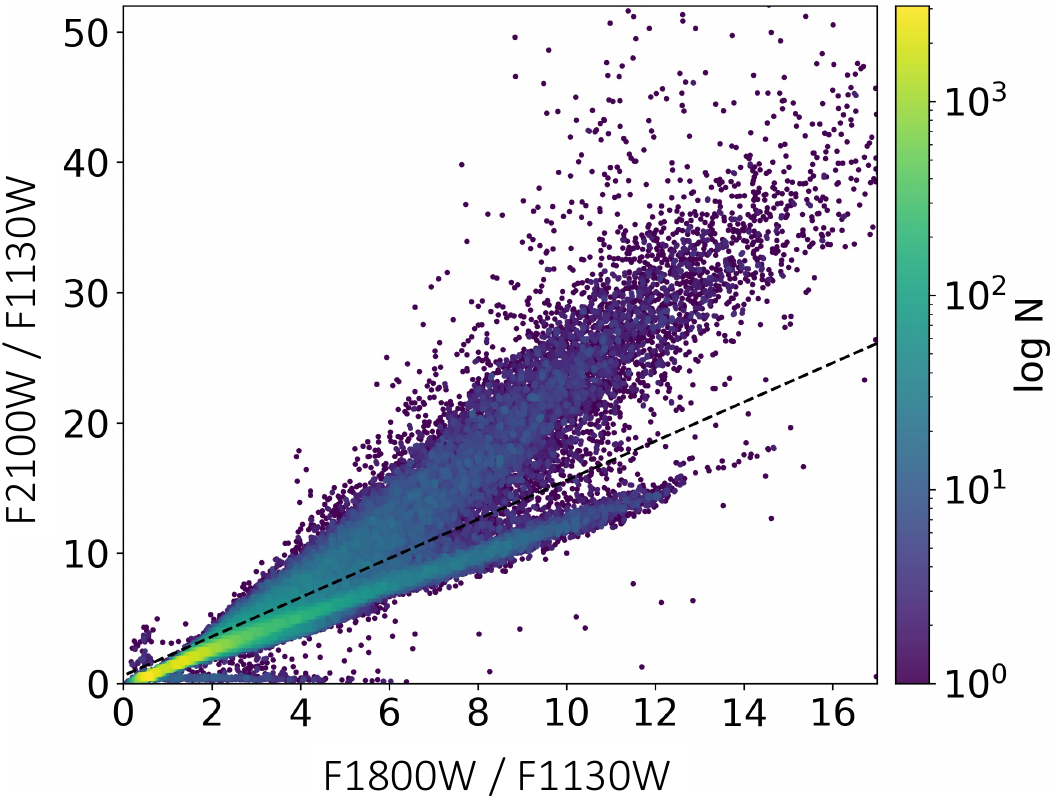}
\caption{\label{colorcolorplot} Plot showing the F2100W-to-F1130W versus the F1800W-to-F1130W flux density ratios for all source pixels in the MIRI mosaics of Cas~A. The colors represent pixel number densities in each region of the plot. The black dashed line (slope of 1.5 and y-intercept of 0.6) separates pixels with a strong F2100W excess in their SEDs (magenta color in Figure~\ref{dust3color}) from those with flatter long-wavelength SEDs.}
\end{figure}

\begin{figure}
\center
\includegraphics[width=0.48\textwidth]{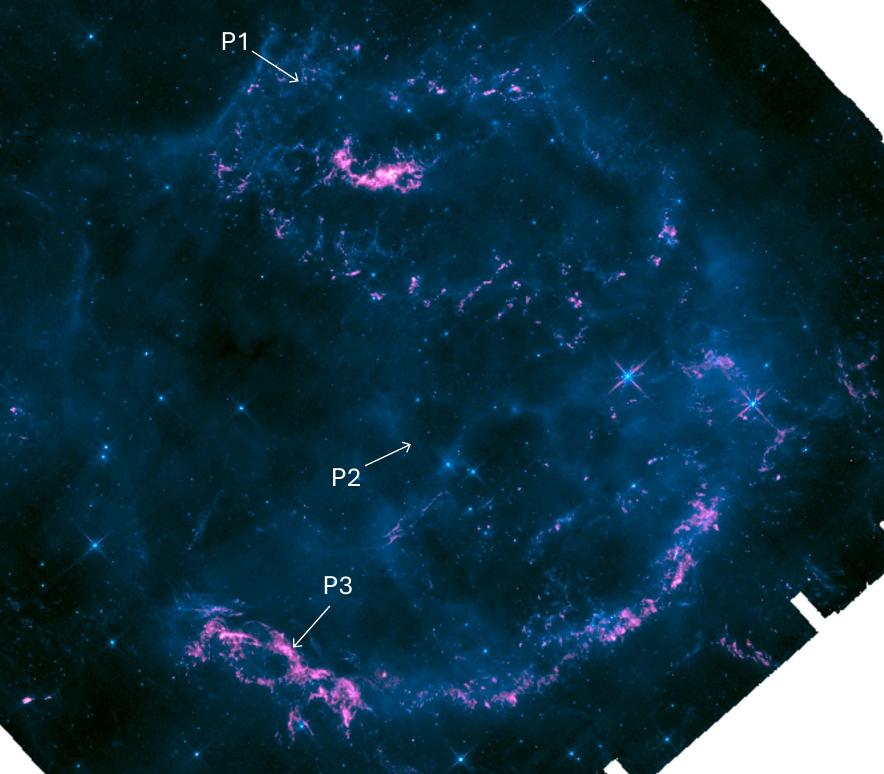}
\caption{\label{carbonimage} Synchrotron-subtracted F560W image, where the significant excess in emission shown in magenta is dominated by emission from carbon dust. The carbon dust emission coincides with regions of enhanced Ne line emission shown as blue regions in Figure~\ref{dust3color}, particularly in the two neon-rich crescents identified by \citet{smith09}. The P1, P2, and P3 regions are indicated by the white arrows. }
\end{figure}

\section{Spatial Distribution of Dust Properties} \label{dustspatial}

Previous \textit{Spitzer} IRS spectral maps of Cas~A showed that the mid-IR dust emission across the remnant can be described by a limited number of characteristic dust types \citep{rho08,arendt14}, although the lower spatial resolution of IRS blended emission from multiple components along individual sightlines. The three MRS positions were intentionally selected to sample regions dominated by these primary dust families, and the higher spatial resolution of JWST has now allowed us to cleanly isolate the individual components. Their distinct MIRI imaging SEDs therefore provide empirical templates that can be used to identify regions across the remnant with similar emission properties and examine the corresponding spatial distributions of the individual dust and gas components.

\subsection{Dust Composition Maps} \label{dustdecomposition}

Using the spectra from the regions sampled by MRS, we computed the resulting MIRI imaging spectral energy distributions (SEDs) for positions P1, P2, and P3. The spectra for the three positions with the MIRI filter bandpasses overlaid are shown in the top panel of Figure~\ref{dustseds}. The bottom panel of Figure~\ref{dustseds} shows the resulting SEDs.

The SED for P1 can be differentiated from the others by a strong excess in the F2100W filter due to the broad 21~$\mu$m feature. The SED at the P2 position is flat across the F1800W and F2100W filters due to the broad bump at 18~$\mu$m. The P3 position shows a relatively featureless gradual rise between 7.7 and 25.5~$\mu$m, but there is a notable excess in the F1280W filter due to the very strong emission from the [\ion{Ne}{2}] line at 12.8~\micron. 
We used these characteristic SEDs to identify pixels across Cas~A with emission properties similar to the three MRS-sampled regions and construct the spatial maps shown in Figure~\ref{dust3color}. We note that this method identifies P1-, P2-, and P3-like SEDs across the SNR, but does not exclude the presence of additional dust compositions in the remnant.

To produce a map of the 21~$\mu$m excess, we created a color-color plot showing the F2100W-to-F1130W versus the F1800W-to-F1130W flux density ratios for all source pixels in the MIRI mosaics of Cas~A, shown in Figure~\ref{colorcolorplot}. The colors in the plot represent pixel number densities, and the black dashed line (slope of 1.5 and y-intercept of 0.6) separates pixels with a strong F2100W excess in their SEDs from those with flatter long-wavelength SEDs. The pixels above this line are those with P1-like SEDs, whose characteristic 21~\micron\ feature is spectroscopically attributed to SiO$_2$. Their spatial distribution is shown in magenta in Figure~\ref{dust3color}.

The pixels below the line in Figure~\ref{colorcolorplot} were further subdivided into those with SEDs that show an F1800W excess and those without. To estimate the F1800W excess, we subtracted the predicted F1800W flux density estimated from a linear interpolation between the F1280W and F2100W flux densities. Pixels with positive residuals were identified as showing excess emission at 18~$\mu$m attributed to Mg$_2$SiO$_4$ CSM emission. This component is shown in green in Figure~\ref{dust3color}. We note that the entire Green Monster structure is identified as  CSM through this method, consistent with the findings of \citet{vink24} and \citet{delooze24}, providing an independent confirmation of the CSM classification.

Within the subset of pixels that do not show an F1800W excess, we selected those that show an excess at F1280W from [\ion{Ne}{2}] emission, as seen in the SED at P3 (Figure~\ref{dustseds}). To estimate the excess, we subtracted the predicted F1280W flux density, obtained via linear interpolation between the F1000W and F1130W flux densities, from these pixels. The pixels with positive residuals are assumed to have a P3-like spectrum with excess neon emission. These regions are shown in blue in Figure~\ref{dust3color}. To independently confirm the association of carbon dust with this same region, we subtracted a scaled synchrotron-dominated F356W NIRCam image from the F560W MIRI image using an average synchrotron spectral index of $-0.68$ \citep{rho24}. Based on the bottom panels of Figure~\ref{dustfits}, the residual emission in the F560W image should be dominated by carbon dust. This synchrotron-subtracted F560W image tracing carbon dust emission is shown in Figure~\ref{carbonimage}. The brightest regions in this image (colored in magenta) closely correlate with the blue neon-bright emission in Figure~\ref{dust3color}, providing additional support that the Ne-rich emission traces the same carbon dust component identified spectroscopically at P3.
The remaining pixels without an obvious excess in the F1280W, F1800W, or F2100W filters are shown as light brown regions in Figure~\ref{dust3color}. These regions tend to be faint and could arise from SN ejecta (with line emission contributing to their SEDs) and/or emission from cooler dust in the CSM that peaks longward of 18~\micron.

\begin{figure*}
\center
\includegraphics[width=0.8\textwidth]{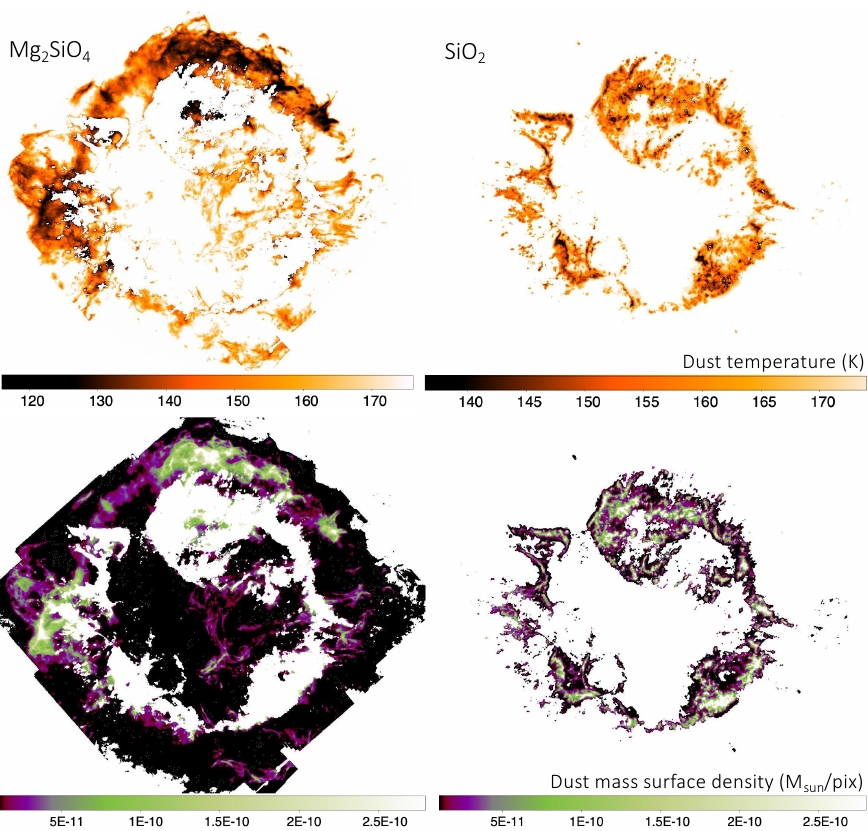}
\caption{\label{dusttempmass} Estimated dust temperature and mass surface density distributions for the CSM (left) and SN ejecta (right) components in Cas~A. The spatial distributions were estimated from a single-temperature fit to the dust-emission-dominated (F1130W, F1800W, and F2100W filters) SEDs of each pixel in the CSM and ejecta regions (see Section~\ref{tm_maps}), and therefore include only the warm, mid-IR-emitting grains whose mass represents a lower limit on the overall dust mass. The fits assume that the emission is dominated by Mg$_2$SiO$_4$ grains in the CSM and SiO$_2$ CDE2 grains in the ejecta. The neon-rich ejecta regions that contain carbon dust are not included.}
\end{figure*}

\subsection{Dust Temperature and Mass Distribution} \label{tm_maps}

Since the P1 and P2 regions can be fairly well approximated by a single average temperature of SiO$_2$ and Mg$_2$SiO$_4$ grains, respectively (see Figure~\ref{dustfits}), we produced average dust temperature and dust mass surface density maps for these compositions. We used the P1- and P2-like spatial templates shown in Figure~\ref{dust3color} and fitted the SEDs for each pixel within those regions with a single-temperature dust model of the corresponding MRS-derived composition.
We only used the F1130W, F1800W, and F2100W MIRI filter flux densities in the SEDs that are dominated by dust emission and do not contain a significant contribution from lines. The resulting dust temperature and mass surface density maps are shown in Figure~\ref{dusttempmass}. These maps reflect the average dust temperature of warm grains emitting in the mid-IR and do not capture the colder grains. The mass surface density therefore does not represent the absolute dust mass, but it does convey the relative dust mass distribution responsible for the mid-IR emission observed with JWST. Integrating the map over the entire SNR provides a lower limit on the dust mass. This value is on the order of $\sim$ 10$^{-4}$~M$_{\odot}$ for each of the SiO$_2$ and Mg$_2$SiO$_4$ components. We did not calculate the maps for the carbon component since the emission could not be well represented by a single dust temperature and the mass estimate is dominated by the poorly constrained minimum temperature.

The temperature distribution for Mg$_2$SiO$_4$ that we associate with the CSM shows that the dust in the northeastern half of the shell is cooler than the southwestern half, which includes the Green Monster region. The northeastern half of the shell also shows a higher dust mass surface density. While this may reflect real differences in the CSM dust mass distribution, it may also be caused by a higher contribution of smaller grains in the southwestern half that would increase the average dust temperature measured in the mid-IR and consequently lower the mass estimate.

The temperature and mass distributions of SiO$_2$ dust associated with the SN ejecta show cooler concentrated clumps of emission embedded in more diffuse warmer dust. This is consistent with the clumpy nature of the ejecta distribution discussed in \citet{milisavljevic24}.

\section{Discussion} \label{disc}

\subsection{Summary of Observational Properties}

JWST MRS observations provided spectra of small-scale emission, allowing us to constrain the dust composition in individual gas clumps with distinct chemistries. In Section~\ref{dustcomposition}, we showed that while we cannot constrain the temperature distribution and the total dust mass in the regions sampled by MRS, we can conclude that dust emission in the mid-IR is dominated by SiO$_2$, Mg$_2$SiO$_4$, and carbon grains at positions P1, P2, and P3, respectively. Dust emission at P2 and P3 is well reproduced by a single dominant grain composition, while emission at P1 requires an additional dust component, with SiO$_2$ providing the primary contribution to the broad 9, 12, and 21~$\mu$m features. These findings are consistent with the prediction from dust condensation models that specific grain species form in chemically distinct nucleosynthetic zones \citep[e.g.,][]{nozawa10, cherchneff26}, and indicate that the multi-component fits required by prior \textit{Spitzer} IRS observations \citep{rho08,arendt14} largely arose from spatial blending of emission from multiple chemically distinct regions rather than a physical coexistence of different grain species within the same ejecta clumps. This has important implications for the interpretation of mid-IR emission from unresolved extragalactic SNe \citep[e.g.,][]{Shahbandeh+2023,zsiro25,Shahbandeh+2025,tinyanont25,sarangi25,szalai25,clayton25}, which likely originates from chemically distinct ejecta and CSM regions containing multiple dust species, temperatures, and heating mechanisms. Further investigation is needed to explore how this complexity impacts the inferred dust compositions and masses in these sources.

The identification of warm Mg$_2$SiO$_4$ grains at P2 provides complementary information about the circumstellar environment into which Cas~A is expanding. The spatial distribution map in Figure~\ref{dust3color} indicates that Mg$_2$SiO$_4$ emission traces a large fraction of the shocked CSM surrounding the remnant, including the Green Monster region \citep{vink24,delooze24}.

Next, we concentrate on positions P1 and P3, which sample the emission from the shocked SN ejecta. SN dust condensation models generally predict that Mg-silicate grains will dominate in ejecta regions where both Mg and Si are present \citep[e.g.,][]{kozasa09, nozawa10, sarangi15,brooker2022}, unless the temperature in the oxygen core is much lower than the standard case \citep{cherchneff26}. The prominence of mid-IR SiO$_2$ emission relative to Mg-bearing silicates does not directly trace their total dust masses and will depend on the relative grain-size distributions \citep[e.g.,][]{marassi2025} and optical properties of each species. Nevertheless, the presence of SiO$_2$ 
favors ejecta that are relatively Si-rich or Mg-poor compared with conditions that would produce Mg-rich silicates.
While reverse-shock processing of small Mg-bearing silicate grains could, in principle, produce Mg-depleted, Si/O-rich grain surfaces, laboratory irradiation experiments show that preferential Mg depletion is energy-dependent and that compositional changes are not observed at the higher ion energies expected in Cas A \citep{demyk2001,jager2003}. We therefore interpret the presence of the SiO$_2$-rich dust component as evidence of ejecta with a relatively high Si/Mg abundance ratio.

The strong 21~\micron\ feature from SiO$_2$ in the P1 spectrum is present throughout much of the reverse-shocked ejecta, consistent with the \textit{Spitzer} IRS spectral maps of Cas~A presented in \citet{arendt14}, and now confirmed at higher spatial resolution by our decomposition map shown in Figure~\ref{dust3color}.
This suggests that conditions favorable to SiO$_2$ dust formation are a widespread property of the reverse-shocked ejecta rather than a localized feature of P1. These regions with a strong 21~\micron\ feature are also found to spatially correlate with particularly strong Ar line emission, offering additional clues to the abundance structure associated with the SiO$_2$-rich ejecta \citep{arendt14}.

The widespread association of SiO$_2$-rich dust with the reverse-shocked ejecta is particularly relevant when compared with previous X-ray abundance measurements. Sato et al. (2025) measured substantial spatial variations in Si/Mg and Ne/Mg across the O-rich ejecta and interpreted the Si-rich/Ne-poor component as evidence for incomplete shell-merger mixing. However, for refractory elements such as Si and Mg, the X-ray line emission traces only the gas-phase component of the shocked ejecta and does not account for material that may have condensed into dust.
Since refractory elements can be efficiently incorporated into dust, as evidenced by their strong gas-phase depletions in the ISM \citep[e.g.][]{jenkins2009} and large dust masses observed in some SNRs \citep[e.g.][]{matsuura11}, the condensed component may represent a significant fraction of the total Si and Mg abundances and must be considered together with the gas-phase abundances when inferring the total ejecta composition.
Our identification of SiO$_2$ as the dominant carrier of the 21~\micron\ feature constrains the composition of the condensed material and provides independent evidence for relatively high Si/Mg in the ejecta.

In contrast, the spectrum at P3 is dominated by carbon dust and shows particularly strong Ne emission relative to P1. The relatively high carbon abundance is supported by the identification of strong CO emission in the same region \citep{rho24}. Since most carbon in SN ejecta is expected to be locked up in CO, the presence of carbon dust suggests that residual carbon was available for dust formation, favoring ejecta abundances with C/O $>$ 1. Figure~\ref{carbonimage} demonstrates that the spatial distribution of carbon dust correlates with the Ne-bright emission in the reverse-shocked ejecta seen in Figure~\ref{dust3color}. 

The Ne emission is heavily concentrated in two opposing crescent-shaped regions, oriented along the inferred kick vector of the compact object in Cas~A \citep{smith09}. These two regions were estimated to contain as much as $\sim$~20\% of the total neon mass in the remnant, with the lower limit on their Ne/Ar abundance ratio estimated to be $\sim$26 \citep{smith09}.

In summary, ejecta emission at P1 is Ar-rich and shows SiO$_2$-dominated mid-IR emission, which we interpret as Si-rich and Mg-poor ejecta conditions, while the carbon-dominated dust at P3 favors carbon-rich ejecta conditions with C/O $>1$ and is associated with a relatively high Ne abundance. The observed properties for the P1 and P3 regions and their inferred ejecta conditions are summarized in Table~\ref{obs_properties}.

\begin{figure}
\center
\includegraphics[width=0.48\textwidth]{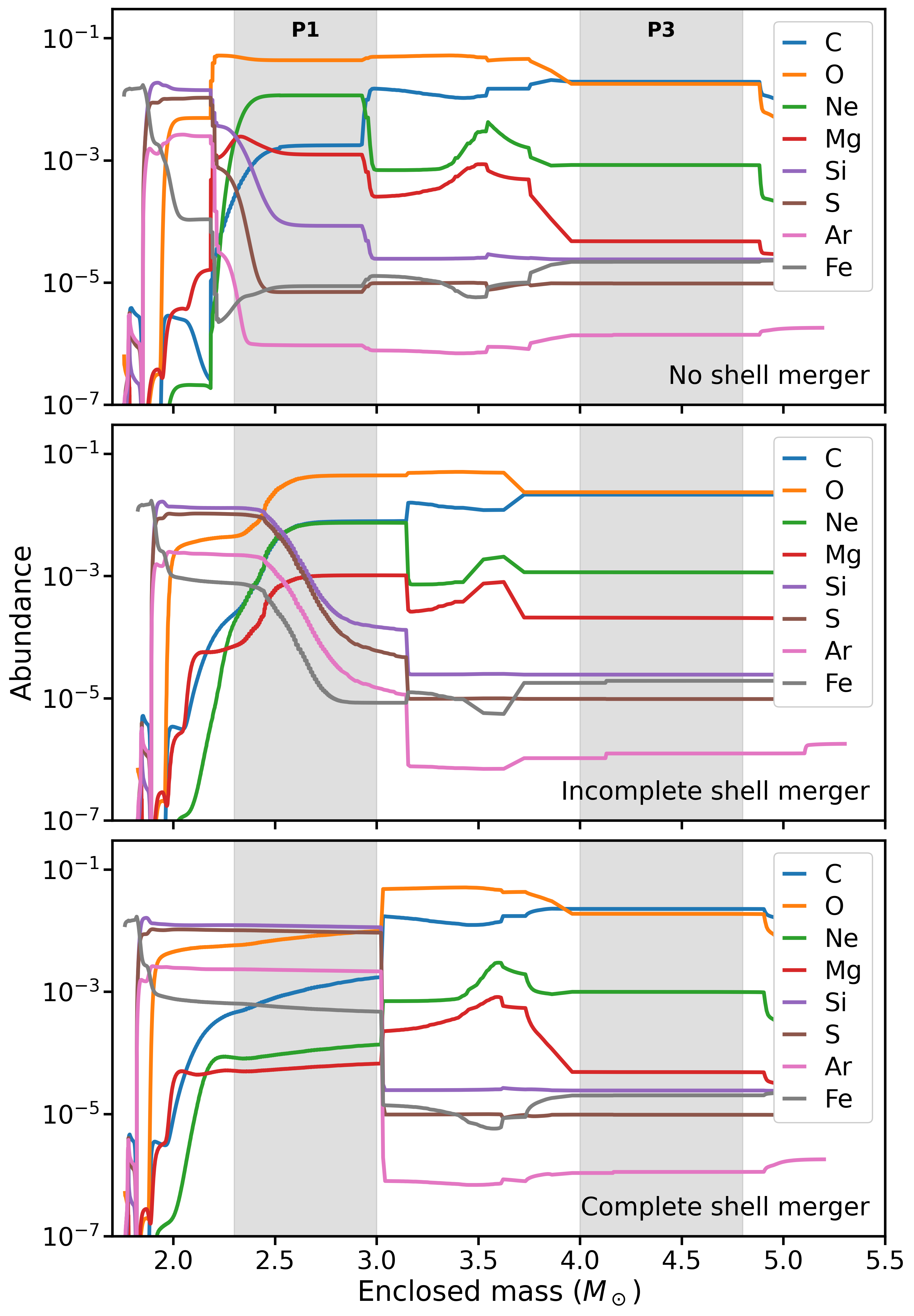} 
\caption{\label{progenitors} Post-explosion ejecta abundances (mass fraction divided by the mass number of the isotope) as a function of the enclosed mass for the three models listed in Table~\ref{sn_models} \citep{boccioli26,boccioli26b}. The top panel is the progenitor with standard oxygen- and neon-rich layers that did not experience a C-O shell merger before the explosion; the middle panel is the progenitor that has undergone a partial/incomplete shell merger; and the bottom panel is the progenitor with a complete shell merger. The gray bands represent our proposed ejecta zones corresponding to the P1 and P3 positions in Cas~A.
}
\end{figure}

\begin{figure*}
\centering
\includegraphics[width=0.4\textwidth]{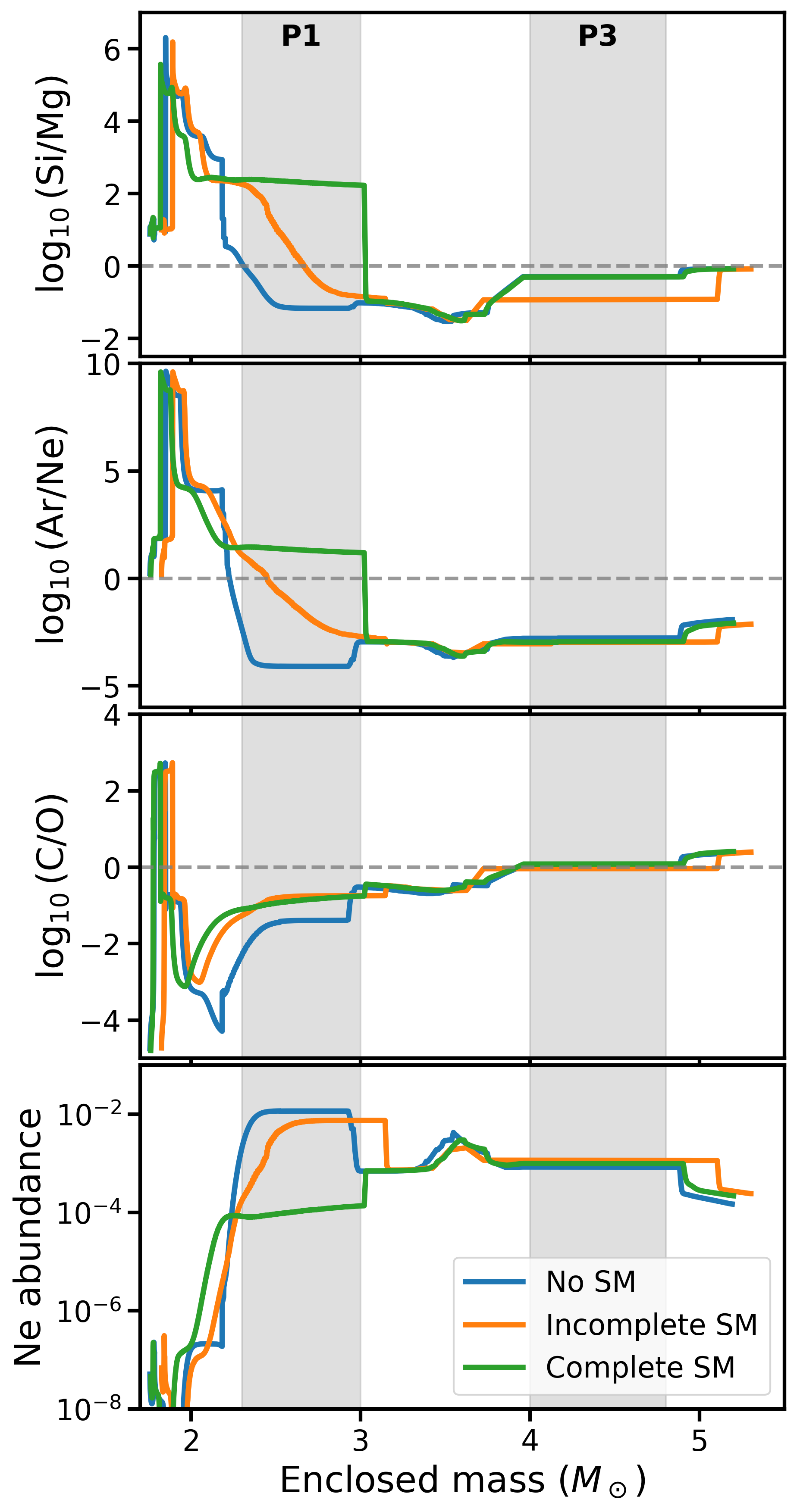}
\includegraphics[width=0.4\textwidth]{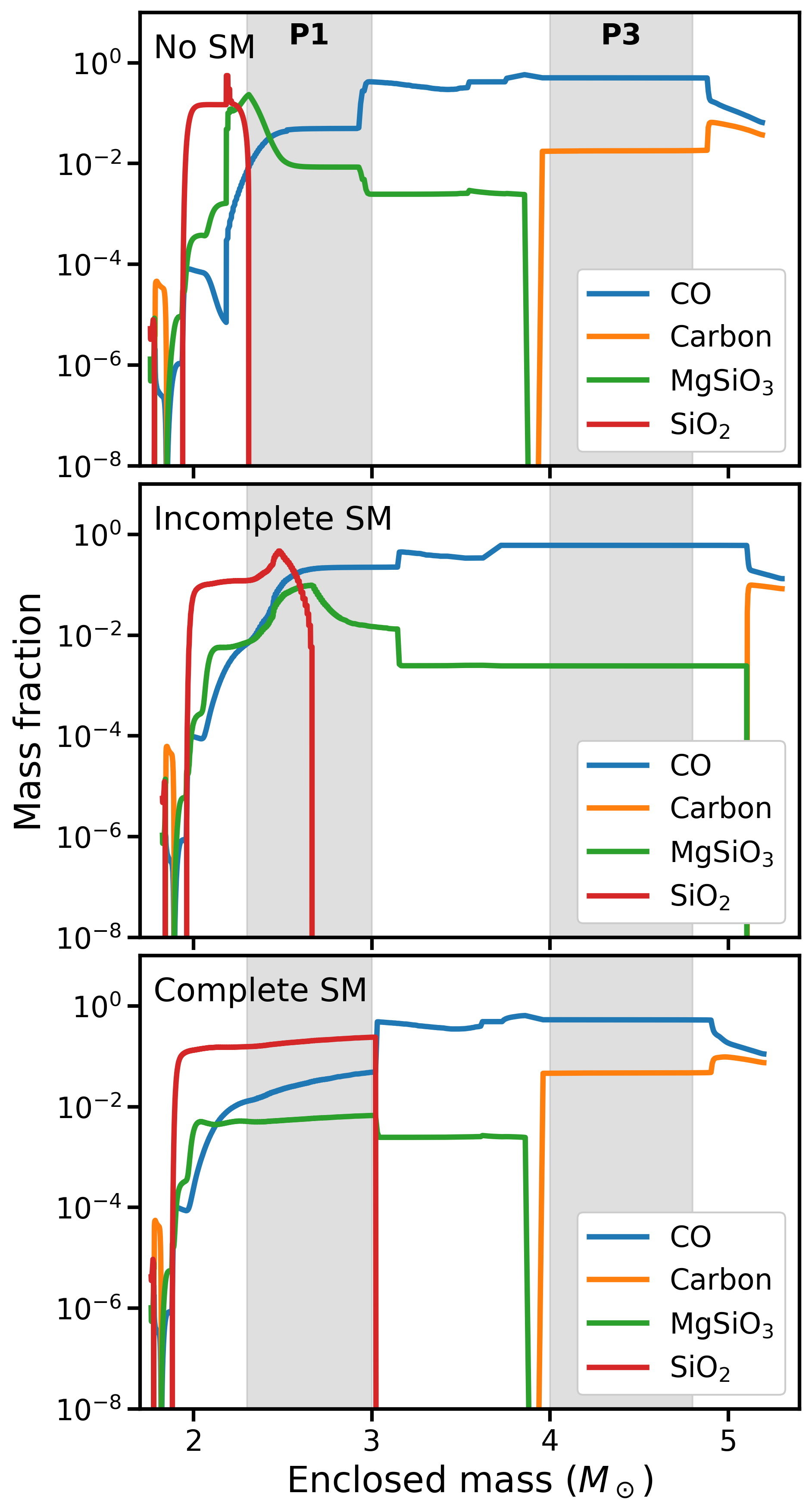}
\caption{\label{gas_diagnostics} \textbf{Left:} Silicon-to-magnesium ratio, argon-to-neon ratio, carbon-to-oxygen ratio, and the neon abundance (neon mass fraction divided by the mass number) as a function of the enclosed stellar mass for the models listed in Table~\ref{sn_models} and yields shown in Figure~\ref{progenitors}. The horizontal dashed lines indicate an element number ratio of unity. In this comparison, Si/Mg $>$ 1 is used as a proxy for ejecta zones favorable to SiO$_2$ formation, rather than as a direct abundance measurement from the observed dust spectrum. The gray bands represent proposed ejecta zones corresponding to the P1 and P3 positions in Cas~A, the same as those shown in Figures~\ref{progenitors}. These zones can simultaneously satisfy the observed properties and inferred abundance conditions summarized in Table~\ref{obs_properties} for the C-O shell merger (SM) model, but not for the standard progenitor model without a merger. \textbf{Right:} Maximum possible mass fraction of CO, carbon, MgSiO$_3$ and SiO$_2$ as a function of enclosed mass for the models listed in Table~\ref{sn_models} and yields shown in Figure~\ref{progenitors}. The values do not represent predictions from condensation models but rather a simple calculation of available yields under the assumption that all spatially coincident O and C first form CO and the remaining carbon atoms form carbon dust. Furthermore, we assume that coincident Si and Mg first form MgSiO$_3$ grains, and residual Si atoms form SiO$_2$. It can be seen that more complete C–O shell mergers (SM) produce larger SiO$_2$ mass fractions and broader spatial regions favorable for SiO$_2$ formation.}
\end{figure*}

\begin{table}
\centering
\setlength{\tabcolsep}{2pt}
\begin{tabular}{cc|ccccl}
& Model ID & $E_{\rm exp}$ & $M^{\rm bary}_{\rm NS}$ & $M_{\rm ej}$ & $M_{\rm pre-SN}$ & Shell  \\
& from B26 & ($10^{51}$~erg) & (M$_{\odot}$) & (M$_{\odot}$) & (M$_{\odot}$) & Merger \\
\hline
1. & 013a200  & 1.002 &  1.756 & 3.434 & 5.191 & None\\
2. & 013a300  & 0.799 &  1.822 & 3.477 & 5.302 & Incomplete \\
3. & 013a250  & 0.816 &  1.734 & 3.464 & 5.200 & Complete \\
\hline
\end{tabular}
\caption{\label{sn_models}Properties of the core-collapse supernova models from \citet{boccioli26} (B26) used in our comparisons to Cas~A.  The model ID designations from \citet{boccioli26} are listed in the first column. The stellar progenitor models are from 13~M$_{\odot}$ ZAMS-mass rotating stars from \citet{limongi18} (013a300) and \citet{falla25} (013a200 and 013a250) that either lack a C-O shell merger or have undergone a partial/incomplete or a complete shell merger prior to explosion (last column). The listed parameters are the SN explosion energy ($E_{\rm exp}$), the baryonic mass of the neutron star ($M^{\rm bary}_{\rm NS}$), i.e. the mass cut, SN ejecta mass ($M_{\rm ej}$), and the pre-SN stellar mass ($M_{\rm pre-SN}$).}
\end{table}


\subsection{Comparison with Explosion Models}

The distinct nucleosynthetic signatures of a shell merger progenitor include enhanced abundances of Si, S, Ar, Ca, and odd-Z elements such as P, Cl, K, and Sc, together with relatively low abundances of Ne and Mg \citep[e.g.,][]{ritter2018}. Evidence for such a progenitor in Cas~A includes the observed inhomogeneous mixing of Si-rich and Ne-rich material \citep{sato2025}, consistent with 3-D simulations of merger progenitors \citep{yadav20}, detection of odd-Z elements of P, Cl, and K in the ejecta \citep{xrism26}, and comparison of explosion nucleosynthesis models for shell merger progenitors with the dust-insensitive Ar/Ne abundance ratio in Cas~A \citep{boccioli26b}. Here, we compare the dust and gas properties described above with explosion nucleosynthesis models to evaluate the shell-merger scenario proposed for Cas~A. 

The properties of dust and gas at positions P1 and P3 (Table~\ref{obs_properties}) are compared with 1-D SN explosion nucleosynthesis models of standard and shell-merger progenitors described in \citet{boccioli26} and \citet{boccioli26b}. We chose to compare explosion nucleosynthesis yields for three different rotating stellar progenitor models with similar zero-age-main-sequence masses of $\sim$ 13~M$_{\odot}$ and stripped hydrogen envelopes, leaving  a total ejecta mass of $\sim$~3.5~M$_{\odot}$ that is broadly consistent with the mass estimated for Cas~A \citep{hwang12,laming20}. The parameters for the SN models are listed in Table~\ref{sn_models}. We note that these models are not intended to reproduce Cas~A specifically, but rather to illustrate the general trends that distinguish the final yields of shell-merger progenitors from those of more standard progenitors. 

The largest difference between the three progenitor models is the degree to which the O and C shells have merged. The progenitors for the first and third models in Table~\ref{sn_models} from \citet{falla25} represent the two most extreme cases, corresponding to no shell merger and a complete shell merger, respectively, while the progenitor used in the second model from \citet{limongi18} experienced a partial or incomplete C-O shell merger \citep[][also private communication]{boccioli26b}. The elemental abundances of the post-explosion ejecta for the three progenitor models as a function of the enclosed mass are shown in Figure~\ref{progenitors}. The most notable difference between the model yields shown in Figure~\ref{progenitors} is the increased extent of the Si-rich region with high Ar/Ne abundance ratios, which develops as a result of the C--O shell mergers. In these models, ingestion of C- and Ne-rich material into the O-burning shell enhances O-burning and activates the associated proton-capture reactions, producing heavy $\alpha$-elements such as Si and Ar while depleting Ne and Mg \citep{roberti25}. These abundance signatures are then expected to be observed in the more extended O-burning shell ejecta, and not just in the deepest layers of ejecta that underwent explosive Si burning.

Using the yields in Figure~\ref{progenitors}, we produced plots of diagnostics that are most relevant to the observed properties and inferred abundance conditions listed in Table~\ref{obs_properties}. The left panel of Figure~\ref{gas_diagnostics} shows plots of the Si/Mg, C/O, and Ar/Ne abundance ratios, as well as the total Ne abundance as a function of enclosed mass. The right panel of Figure~\ref{gas_diagnostics} shows the maximum possible mass fraction that could be locked up in CO molecules and carbon, MgSiO$_3$, and SiO$_2$ dust grains, purely based on the available elemental yields for each mass zone. The values were calculated under the assumption that all spatially coinciding O and C first form CO, and the remaining carbon forms carbon dust. We also assume that coincident Si and Mg first form MgSiO$_3$ grains, while the remaining Si forms SiO$_2$. We stress that these dust mass fractions are not based on condensation models and should not be used to estimate expected dust masses. They are nevertheless informative about where in the ejecta the observed dust species could possibly form. Due to the increasing extent of the Si-zone, the maximum possible mass fraction of SiO$_2$ dust grows significantly for shell merger explosions, as does the total dust mass. 

Using the conditions summarized in Table~\ref{obs_properties}, we qualitatively considered which model could simultaneously produce an ejecta region characterized by conditions favorable to SiO$_2$ dust formation and elevated Ar abundances, and another characterized by carbon dust and a high abundance of Ne relative to the other region. In the model comparison, we adopt Si/Mg $>$ 1 as an abundance condition favorable to SiO$_2$ formation, and C/O $>$ 1 as a condition favorable to carbon dust formation after the available oxygen is locked up in CO.

Among the models we considered, only the shell-merger models simultaneously satisfy all four conditions summarized in Table~\ref{obs_properties}.
Our proposed ejecta regions that produce the emission at positions P1 and P3 are shown as gray bands in Figures~\ref{progenitors} and \ref{gas_diagnostics}. 
While the inferred conditions at P1 could be produced in the innermost Si-burning ejecta layers of the standard model without a shell merger, the bulk of those layers are likely interior to the material in the reverse-shocked shell. The estimated mass of unshocked ejecta interior to the reverse shock in Cas~A is $\sim$0.5~M$_{\odot}$ \citep{laming20,priestly22}, which, when added to the baryonic neutron-star mass in Table~\ref{sn_models}, gives a lower limit of $\sim$2.3~M${\odot}$ for the enclosed mass of the P1 ejecta zone. 

Table~\ref{obs_properties} summarizes which conditions are met by the standard model without a shell merger versus the model with a complete shell merger. The shell-merger models produce ejecta regions with properties consistent with those inferred for the P1 and P3 regions, particularly extended Si-rich regions with high Ar/Ne abundance ratios. We recognize that there are many caveats in directly comparing 1-D explosion models with observations of Cas~A, since 3-D mixing effects and explosion asymmetries would significantly alter both the progenitor structure and the final ejecta distribution. The distributions of the P1-, P2-, and P3-like emission shown in Figure~\ref{dust3color}, particularly the more localized distribution of P3-like carbon- and Ne-rich emission relative to the P1-like SiO$_2$ emission, are consistent with inhomogeneities that may be typical in shell-merger explosions \citep{yadav20,sato2025}, and illustrate the importance of multidimensional effects in shaping the observed ejecta structure.  Comparison of both the abundance structure and the ejecta morphology with future 3-D explosion models will be needed to further evaluate the shell-merger scenario.

\section{Conclusion} \label{conclusion}

We present an analysis of dust emission spectra obtained with the MIRI MRS at three positions in Cas~A \citep{milisavljevic24} sampling reverse-shocked SN ejecta and the CSM. The high spatial resolution of JWST enabled us to isolate small-scale ejecta clumps with distinct abundance properties and determine the composition of the dust associated with each region. We find that the dust emission in the Ar-bright ejecta region P1 is dominated by SiO$_2$ dust, while the Ne-bright and CO-rich \citep{rho24} ejecta region P3 is dominated by carbon dust. The spectrum from the P2 region within the Green Monster structure associated with the CSM \citep{vink24,delooze24} is well reproduced by warm Mg$_2$SiO$_4$ dust. 
The spatially resolved spectra reveal that individually isolated ejecta clumps are dominated by a small number of grain species, with P2 and P3 well reproduced by single dominant grain compositions and P1 by a Si-bearing dust mixture dominated by SiO$_2$ emission. This supports dust condensation model predictions that specific grain species form in chemically distinct nucleosynthetic zones \citep[e.g.,][]{nozawa10, cherchneff26}, and demonstrates that the multi-component fits required by prior \textit{Spitzer} IRS observations \citep{rho08, arendt14} largely arose from spatial blending along the line of sight rather than physical coexistence of several unrelated grain species within individual knots.
Our dust fits also highlight the importance of accounting for line-of-sight extinction when modeling dust emission from Galactic SNRs in the mid-IR.

Using the characteristic SEDs derived from the MRS regions as empirical templates, we identified regions across the SNR with emission properties similar to the spectroscopically characterized P1, P2, and P3 dust compositions. The P1-like 21~\micron\ emission associated with SiO$_2$ dust is present throughout much of the reverse-shocked ejecta emitting in the IR, while the P3-like Ne-rich emission associated with carbon dust is more localized and concentrated in two crescent-shaped regions first identified by \citet{smith09}. The P2-like emission associated with Mg$_2$SiO$_4$ traces much of the CSM, including the Green Monster structure. The spatial distribution of the different emission templates further demonstrates that the ejecta structure in Cas~A is highly inhomogeneous.

Comparison with 1-D core-collapse SN explosion nucleosynthesis models of \citet{boccioli26b} suggests that the combined gas and dust properties inferred from the observations are more naturally reproduced by progenitors that underwent C--O shell mergers than by standard progenitors without mergers. Spatially resolved dust composition provides an independent constraint on the ejecta abundance structure, particularly for refractory elements such as Si and Mg that are not fully represented by gas-phase measurements. In particular, only the shell-merger models simultaneously produce ejecta zones that are Ar-rich and favor SiO$_2$ dust production, as well as zones with relatively high Ne abundances that favor carbon dust formation. We note that the predictions depend on results from 1-D simulations and caution that comparisons with future 3-D explosion models of shell-merger progenitors will be required to further evaluate the shell-merger scenario. Extension of this analysis to additional MRS positions will also be important to confirm that the identified dust compositions are representative of the ejecta as a whole, particularly in regions whose imaging SEDs do not clearly match the emission from any of the MRS-sampled clumps.

Our study highlights the importance of high-resolution IR observations of SNRs for understanding dust formation in SN ejecta and for interpreting mid-IR dust emission from extragalactic SNe, where multiple chemically distinct regions are spatially unresolved. It also demonstrates that spatially resolved dust compositions in young SNRs can serve as a powerful probe of the chemical structure and explosion properties of core-collapse SNe.

\begin{acknowledgments}

We thank Luca Boccioli for providing the explosion nucleosynthesis models and for his guidance on the model selection.
This work is based on observations made with the NASA/ESA/CSA James Webb Space Telescope. The data presented in this article were obtained from the Mikulski Archive for Space Telescopes (MAST) at the Space Telescope Science Institute, which is operated by the Association of Universities for Research in Astronomy, Inc., under NASA contract NAS 5-03127 for JWST. The specific observations are associated with program 1947 and can be accessed via \dataset[doi:10.17909/yqf7-9a04]{https://doi.org/10.17909/yqf7-9a04}.
The support for the program was provided by NASA grant JWST-GO-01947. T.T. also acknowledges support from NSF grant AST-2205314, and the NASA ADAP award 80NSSC23K1130. Work by R.G.A. was supported by NASA under award number 80GSFC24M0006. D.M. acknowledges support from NSF through grant AST-2206532. J.M.L. was supported by basic research funds of the Office of Naval Research.
I.D.L. acknowledges funding from the European Research Council (ERC) under the European Union's Horizon 2020 research and innovation program DustOrigin (ERC-2019- StG-851622) and from the Belgian Science Policy Office (BELSPO) through the PRODEX project “JWST/MIRI Science exploitation” (C4000142239).

\end{acknowledgments}

\appendix

\setcounter{figure}{0}
\setcounter{table}{0}
\renewcommand{\thefigure}{A\arabic{figure}}
\renewcommand{\thetable}{A\arabic{table}}

\section{Alternative Dust Composition Fits for Position P2}

In Section~\ref{p2dust}, we showed that Mg$_2$SiO$_4$ provides the best overall fit to the MRS spectrum at position P2. To assess the uniqueness of this result, we also fitted the spectrum using several additional grain compositions that have previously been proposed for Cas~A dust or are commonly considered in studies of oxygen-rich CSM dust \citep[e.g.,][]{jones23,kavanagh2026}. These include MgSiO$_3$, Mg$_{2.4}$SiO$_{4.4}$, and MgFeSiO$_4$ from \citet{jager03}, silicates from \citet{ossenkopf1992}, and astrodust from \citet{hensley23}. The Mg$_2$SiO$_4$  fit results are included for completeness. The models were fitted using the same fitting procedure and assumptions described in Section~\ref{p2dust}.

\begin{figure*}
\center
\includegraphics[width=0.46\textwidth]{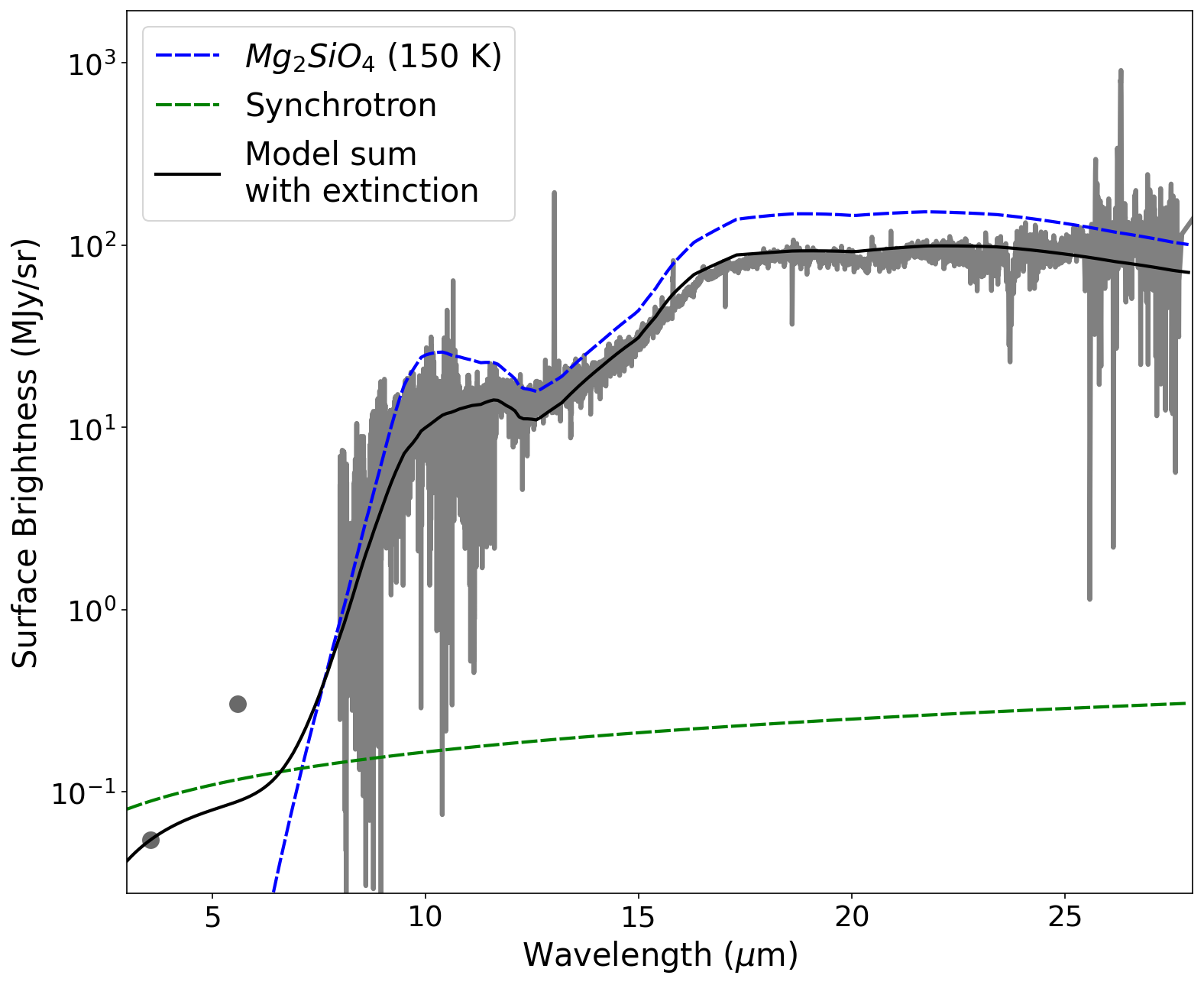}
\includegraphics[width=0.46\textwidth]{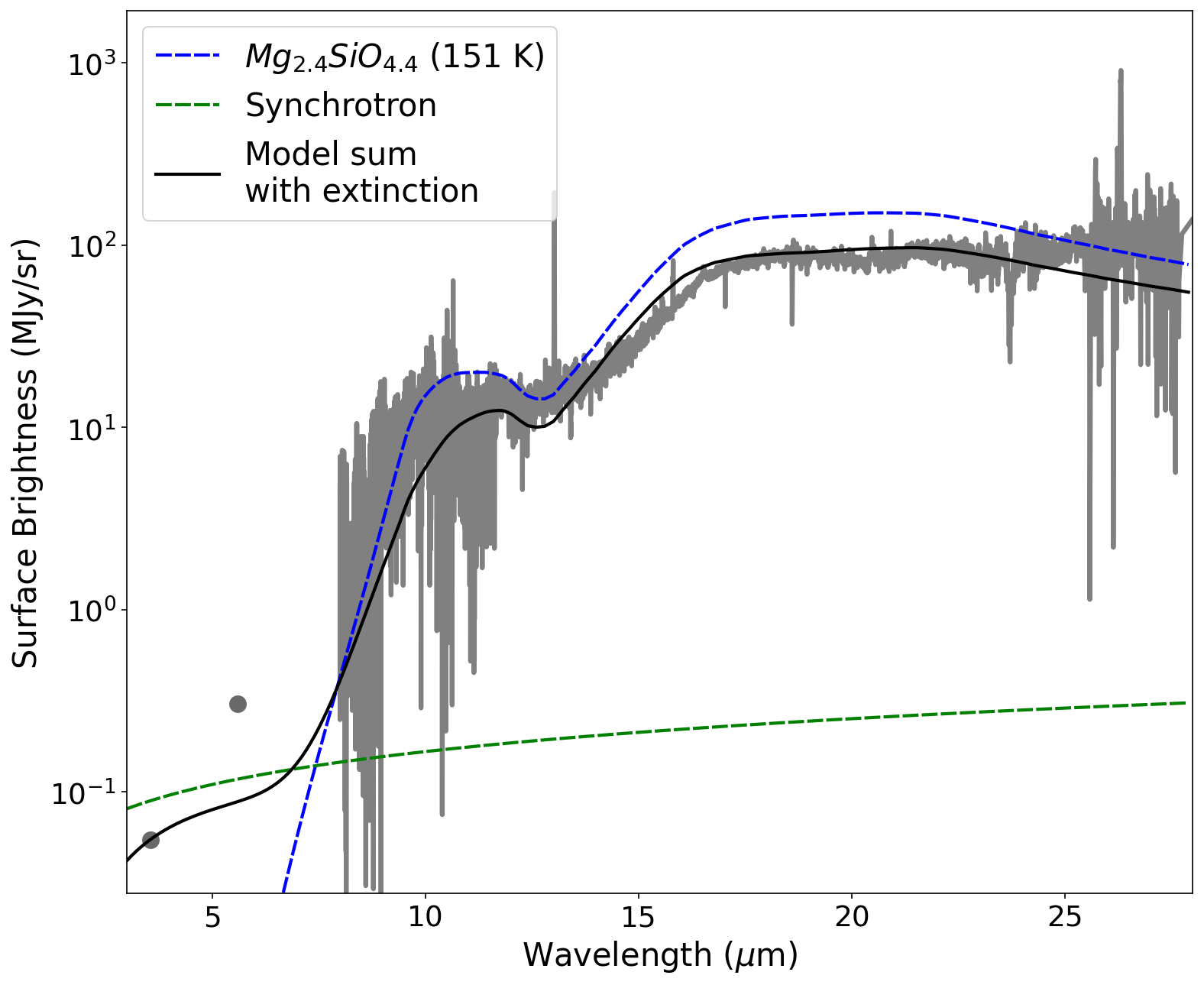}\\
\vspace{-4mm}
\includegraphics[width=0.46\textwidth]{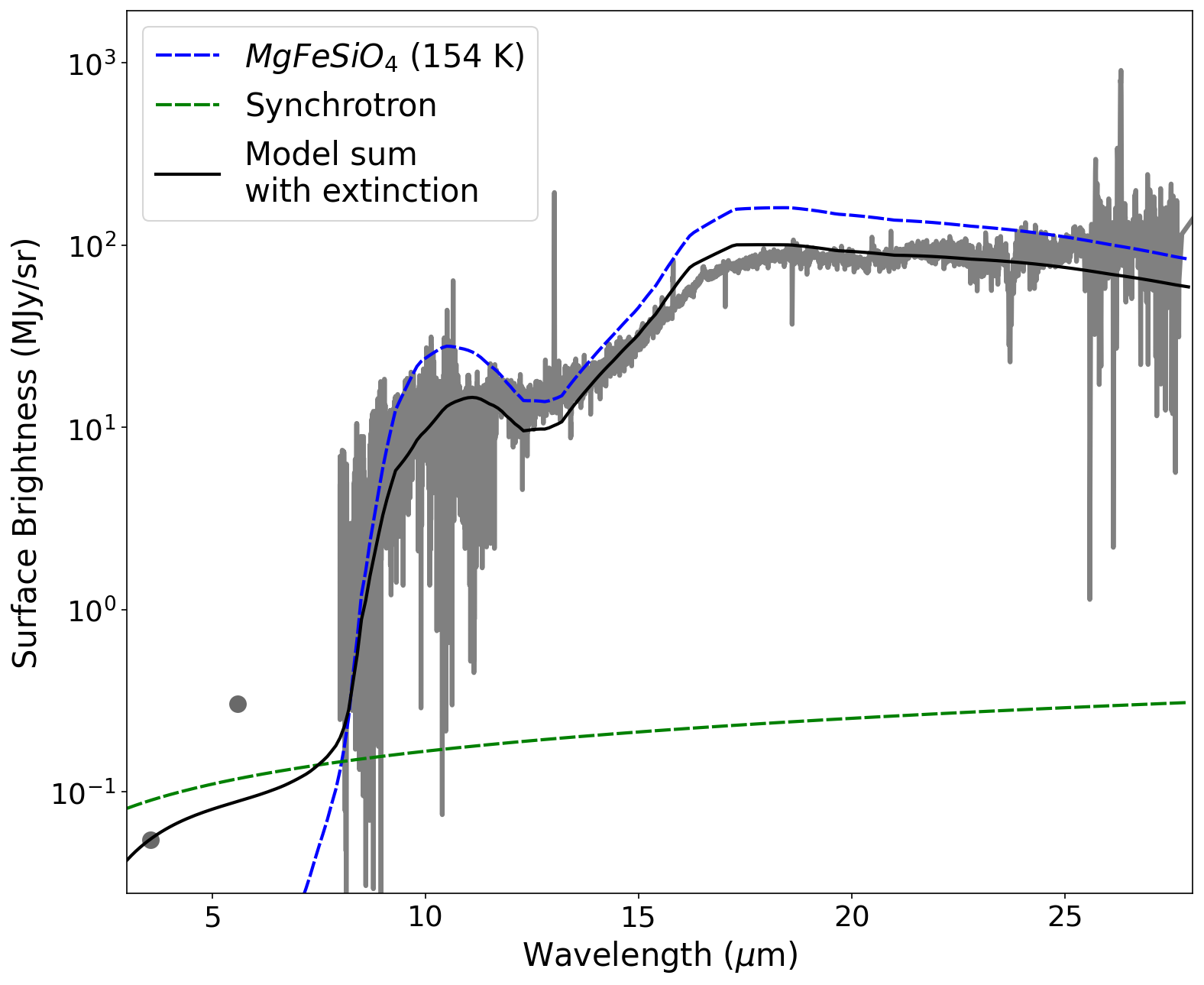}
\includegraphics[width=0.46\textwidth]{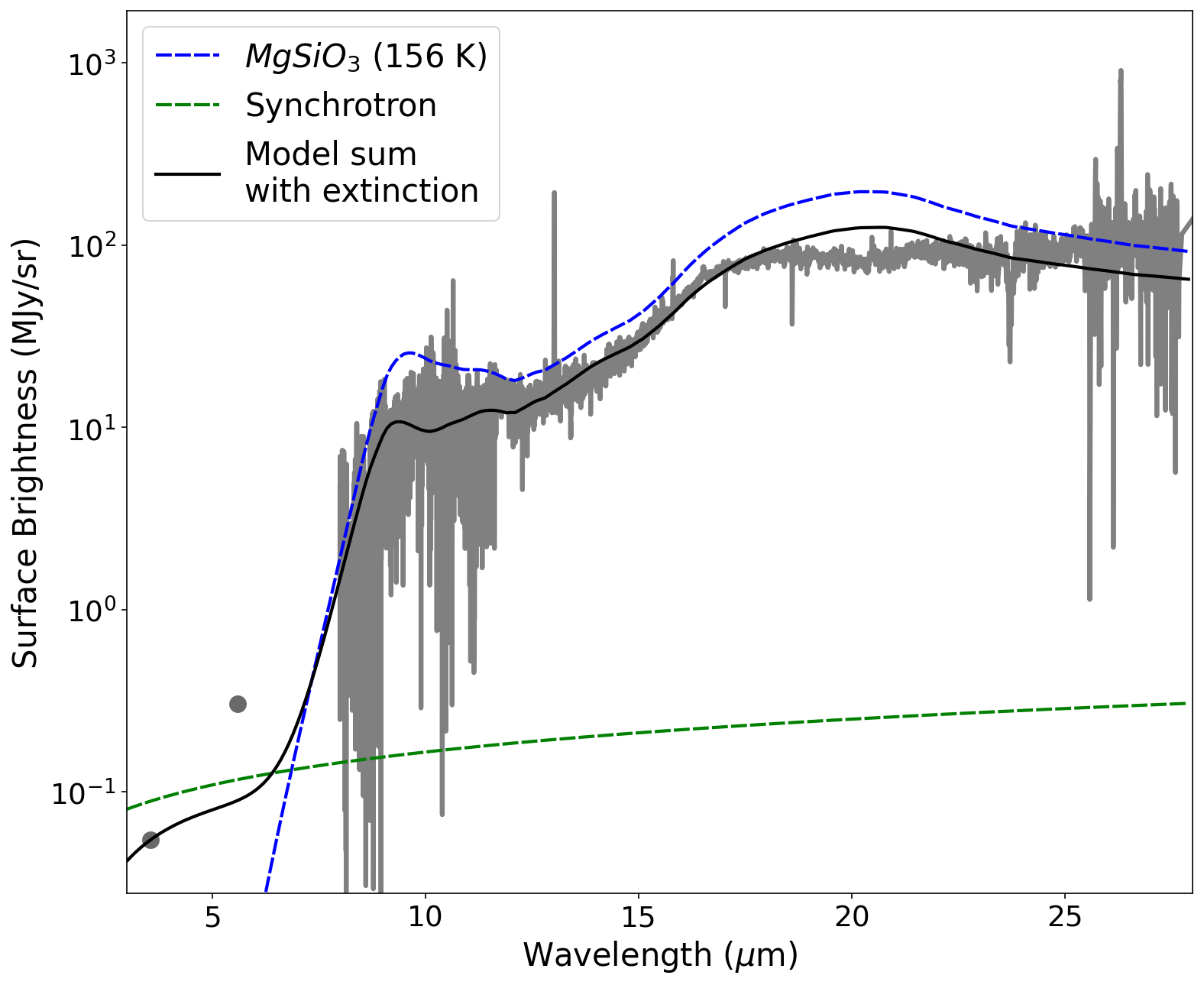}\\
\vspace{-4mm}
\includegraphics[width=0.46\textwidth]{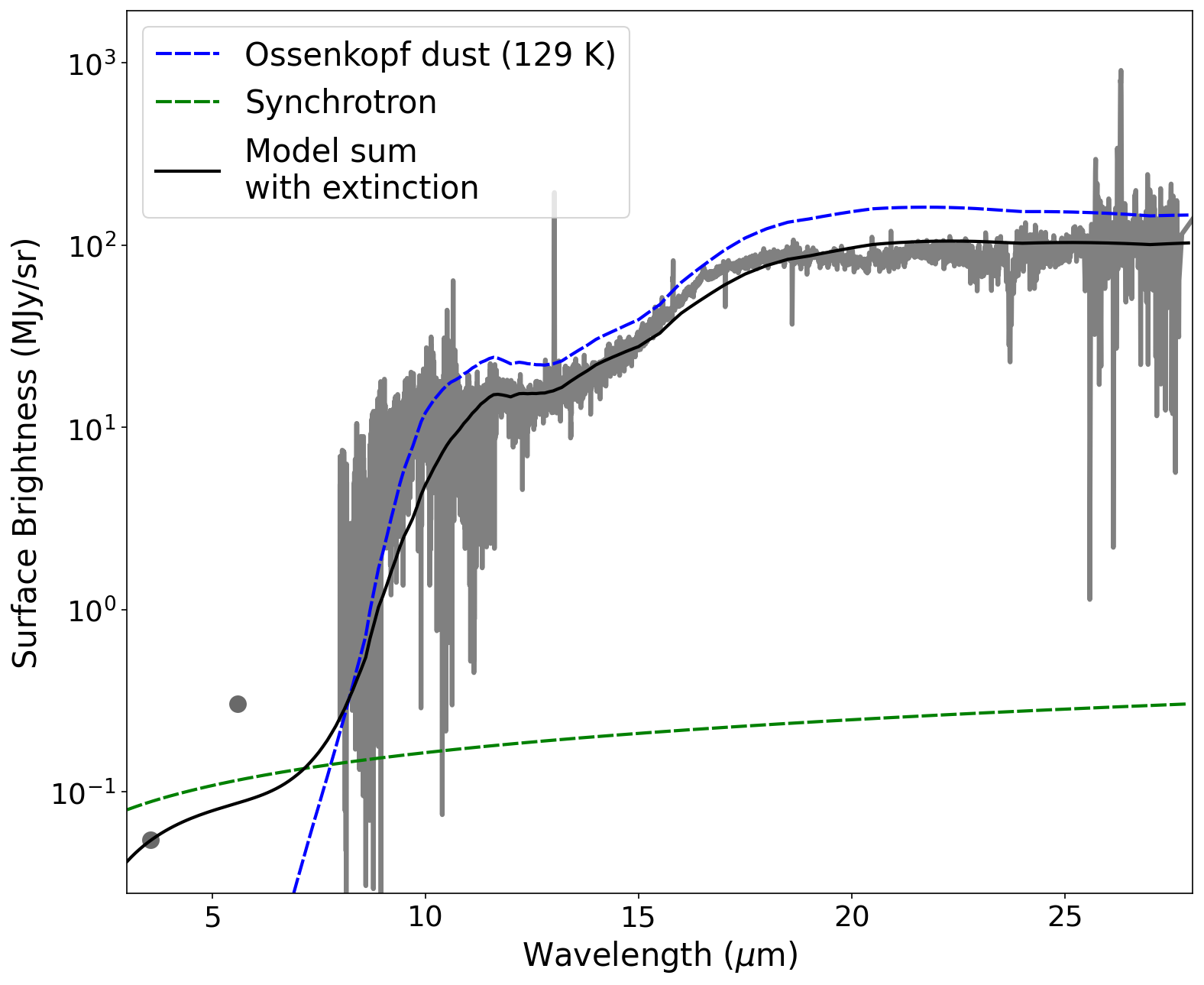}
\includegraphics[width=0.46\textwidth]{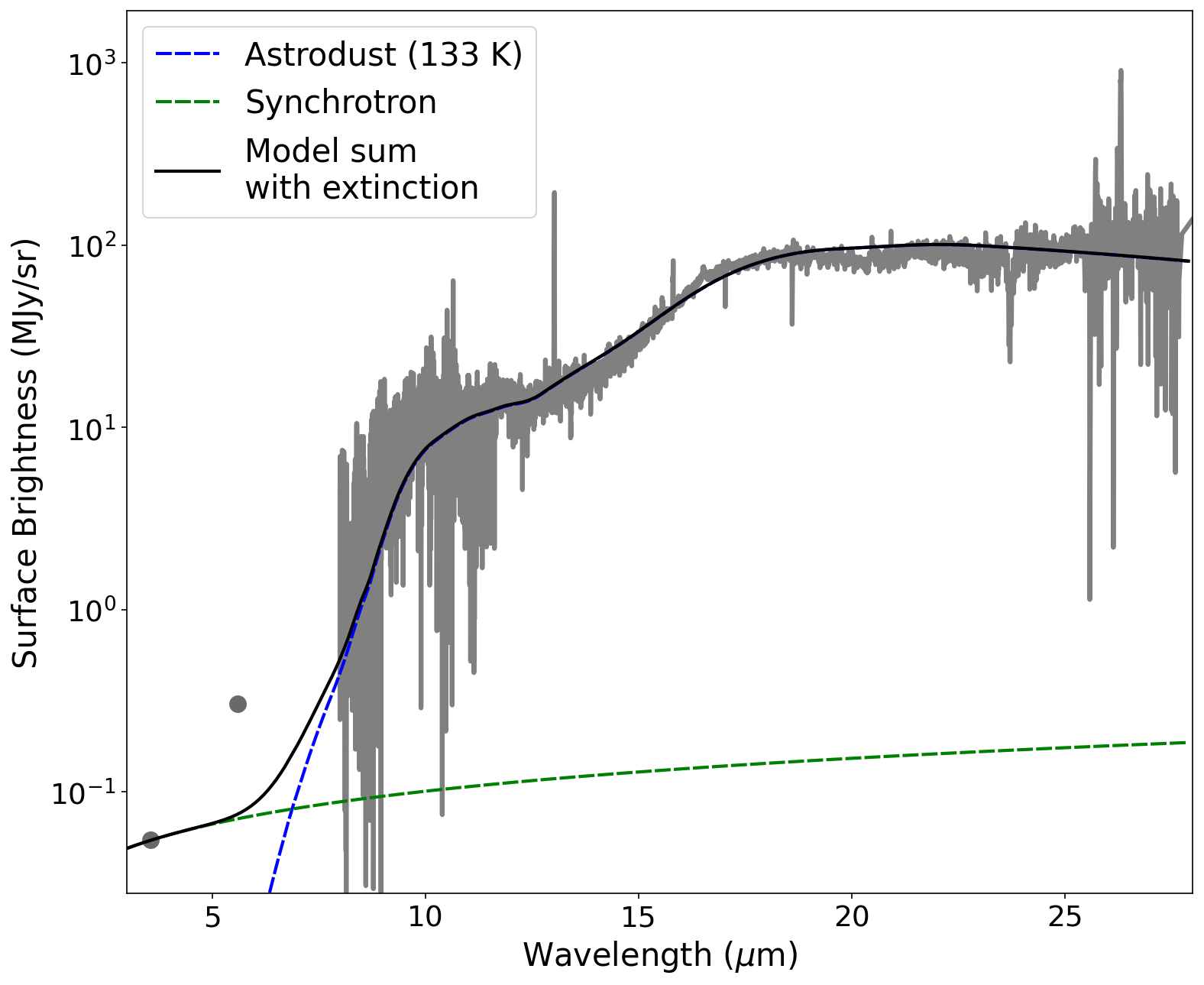}\\
\caption{\label{p2dustfits} Spectral fits to the P2 region using Mg-bearing silicate grain compositions with optical constants from \citet{jager03}, silicates from \citet{ossenkopf1992}, and astrodust from \citet{hensley23}. The fits use a single-temperature dust component with foreground extinction. Gray circles represent the F356W and F560W flux densities from the same extraction apertures, while the green dashed curve represents synchrotron emission with an assumed power-law spectral index of -0.6 \citep{domcek21}. The best-fit parameters for each composition are listed in Table~\ref{tabp2dustfits}. The Mg$_2$SiO$_4$ grains provide the best fit to the spectrum with line-of-sight extinction in the range measured by \citet{delooze17}.
}
\end{figure*}

Figure~\ref{p2dustfits} compares the resulting best-fit spectra, while Table~\ref{tabp2dustfits} summarizes the corresponding best-fit parameters. The uncertainties in the total dust mass and line-of-sight extinction A$_V$ are 1-$\sigma$ statistical uncertainties from the fit. Although several compositions reproduce portions of the spectrum reasonably well, Mg$_2$SiO$_4$ provides the best overall agreement with the broad 18~$\mu$m emission feature while requiring an A$_V$ value in the range measured for Cas~A \citep{domcek21, delooze17}. Although astrodust provides a lower reduced $\chi^2$, it does not reproduce the shape of the 18~$\micron$ feature as well as Mg$_2$SiO$_4$ and it requires an implausibly low A$_V$ value to fit the spectrum.

\begin{table}
\centering
\begin{tabular}{lcccc}
\hline
Composition & Temperature & Dust Mass & $A_V$ & $\chi^2_\nu$ \\
            & (K) & ($10^{-9}\:M_\odot$) & (mag) &  \\
\hline
Mg$_2$SiO$_4$       & $150.4 \pm 0.3$ & $9.6 \pm 0.2$ & $>$ 11.9 & $3.35$ \\
Mg$_{2.4}$SiO$_{4.4}$ & $150.7 \pm 0.4$ & $7.2 \pm 0.2$ & $>$ 11.8  & $4.87$ \\
MgFeSiO$_4$         & $154.3 \pm 0.4$ & $5.6 \pm 0.1$ & $>$ 11.9  & $5.73$ \\
MgSiO$_3$           & $155.5 \pm 0.4$ & $8.6 \pm 0.2$ & $>$ 11.0  & $4.35$ \\
Ossenkopf      & $128.9 \pm 0.2$ & $2.5 \pm 0.1$ & $>$ 11.8 & $4.38$ \\
Astrodust           & $133.3 \pm 0.2$ & $6.5 \pm 0.1$ & $<$ 0.03  & $2.91$ \\
\hline
\end{tabular}
\caption{\label{tabp2dustfits} Best-fit dust model parameters for region P2, assuming compositions of different Mg-bearing silicates from \citet{jager03}, silicates from \citet{ossenkopf1992}, and astrodust from \citet{hensley23}. The quoted uncertainties are 1-$\sigma$ statistical uncertainties from the fit to each grain composition.}
\end{table}

\clearpage

\bibliography{Astro_BIB_new}{}
\bibliographystyle{aasjournalv7.1}

\end{document}